\documentclass[a4paper,11pt]{article}
\usepackage{jcappub} 

\pdfoutput=1

\usepackage{amsmath, latexsym, amssymb, graphicx, ifpdf, slashed, color, hyperref, url, cancel}
\usepackage[T1]{fontenc}
\usepackage[dvipsnames]{xcolor}
\usepackage{enumitem} 
\usepackage{orcidlink}
\usepackage{comment}
\hypersetup{colorlinks, citecolor=nicegreen,linkcolor=nicered, urlcolor=niceblue}
\definecolor{nicered}{rgb}{.7,.1,.1}
\definecolor{nicegreen}{rgb}{.2,.7,.1}
\definecolor{niceblue}{rgb}{0.1,0.2,0.6}
\definecolor{darkblue}{rgb}{0,0,.5}

\newcommand{\Msun}{\, M_\odot}
\newcommand{\MBH}{M_\text{BH}}

\def\Loyola{Department of Physics, Loyola University Chicago, Chicago, IL 60660, USA}
\def\NU{Northwestern University, Department of Physics and Astronomy, Evanston, IL 60208, USA}
\def\Fermilab{Astrophysics Theory Department, Theory Division, Fermi National Accelerator Laboratory, Batavia, Illinois 60510, USA}
\def\adelaide{ARC Centre of Excellence for Dark Matter Particle Physics, Department of Physics, University of Adelaide, South Australia 5005, Australia}

\title{Heavy dark matter in rapidly evolving massive stars}

\author[a,b]{Sandra Robles\,\orcidlink{0000-0002-6046-8217},}
\affiliation[a]{\Fermilab}
\affiliation[b]{Kavli Institute for Cosmological Physics, University of Chicago, Chicago, IL 60637, USA}
\emailAdd{srobles@fnal.gov}

\author[c,d]{Walter~Tangarife\,\orcidlink{0000-0002-4808-3277},}
\affiliation[c]{\Loyola}
\affiliation[d]{\NU}
\emailAdd{wtangarife@luc.edu}

\author[e]{and Giorgio~Busoni\,\orcidlink{0000-0002-8527-0768}}
\affiliation[e]{\adelaide}
\emailAdd{giorgio.busoni@adelaide.edu.au}

\abstract{
We study the impact of heavy dark matter (DM) captured in massive stars via scattering(s) with the star constituents. We focus on the first stars and use stellar evolution simulations to track down how DM capture evolves over time from the zero-age main sequence to the late metal-rich stages of stellar evolution. 
During the early hydrogen–helium–dominated phase, the capture process is well described by scattering with two targets. As a star evolves, metal production leads to the formation of a dense core surrounded by a lighter envelope. The core significantly enhances the capture of ultra-heavy DM; in this case, three distinct nuclear species are required to accurately describe multiple-scattering capture. We use the Eddington inversion method to obtain a \emph{realistic} DM velocity distribution, better suited when the star is near the center of a halo, than the widely used Maxwell-Boltzmann distribution. 
We find that heavy DM would be able to thermalize and achieve capture-annihilation equilibrium within a massive star's lifetime for regions of the parameter space not excluded by direct detection. For non-annihilating DM, because of the high amount of targets available for capture and despite massive stars being short-lived, it would even be possible for DM to achieve self-gravitation and collapse to a black hole, which eventually could swallow the star from within before the expected end of the star's life, for non-excluded regions of the parameter space.  
Our results highlight the dependence of DM capture on the stellar evolutionary stage, composition, and halo location, demonstrating that accurate modeling of massive stars is essential for constraining heavy DM with primordial stellar populations.
}

\begin{document}
\hfill FERMILAB-PUB-25-0962-T


\maketitle

\section{Introduction}

The capture of dark matter (DM) in stars and compact stellar objects has been widely considered in the literature. Seminal work in the 1980s studied capture after one collision between the DM particle and nuclei in the 
Sun and in the Earth~\cite{Press:1985ug,Gould:1987ir},
and later capture after multiple collisions in the Earth's core was considered~\cite{Gould:1991va}. 
More recent studies have extended Gould's original formalism for multiple scattering capture~\cite{Gould:1991va} and applied it to stars and stellar remnants~\cite{Busoni:2017mhe,Bramante:2017xlb,Dasgupta:2019juq,Ilie:2020vec,Leane:2023woh}. However, these studies have assumed that the escape velocity and density are constant throughout the object and that the DM particles follow straight trajectories, unaffected by collisions along the way. These assumptions have been removed in an improved analytical treatment of multi-scattering capture in white dwarfs~\cite{Bell:2024qmj}, in which the escape velocity and star density are functions of the radial position, and gravitational focusing is included. In this analysis, the capture via multiple collisions is encoded in a response function that also accounts for the star's opacity.

In this paper, we tackle the capture of DM in massive stars that rapidly evolve, and for which, as in white dwarfs, infalling DM particles are accelerated to speeds much greater than the DM halo velocity. As a case study, we revisit the capture of DM in the first-generation stars, also known as Population III (Pop.~III) stars. Pop.~III stars formed from metal-free primordial gas in $\sim10^5\,M_\odot\, -\,10^6\, M_\odot$ halos~\cite{Bromm_2002,Klessen:2023qmc}. The formation rate of these stars typically peaks at redshift $z\sim 15-20$; however, they continue to evolve until later cosmic times, producing metals in their cores.  This places Pop.~III stars in a crucial role in connecting metal-free and high-metallicity stars~\cite{Chowdhury:2024wvm}. Since Pop.~III stars are believed to play a crucial role in cosmic reionization~\cite{Bromm:2009uk}, they have been of increasing interest in the literature, especially in the era of JWST observations~\cite{2023AJ_165_2L,Visbal:2025xmd}. 

The capture of DM in Pop.~III stars has been previously explored in both single-scattering and multi-scattering capture regimes, under the above-mentioned assumptions~\cite{Iocco:2008xb,Freese:2008ur,Iocco:2008rb,Yoon:2008km,Ilie:2019sjk,Ilie:2020nzp,Ilie:2021iyh}. In our analysis, we treat the capture of DM in Pop.~III stars using the formalism introduced in Ref.~\cite {Bell:2024qmj} and extending it to the case of multiple scattering with hydrogen, as well as to DM capture by scattering with three different targets, which we found to be relevant for DM capture in the red giant phase. Rather than assuming constant stellar density and escape velocity, we use a series of radial profiles of these quantities obtained from stellar evolution simulations. We calculate the capture rate of DM for masses from $10^3$ GeV to $10^{15}$ GeV at different stages of the evolution of the Pop.~III star. We find relevant differences in the capture rate as the star evolves. The primary driver of this difference is the change in stellar composition as metals are produced. In the late stages of the star, the presence of a metallic core is an essential factor that enhances the capture rate at small DM-nucleon cross sections and must be considered when calculating DM capture. 

Since DM-capture effects are more pronounced in DM-dense environments, we focus on Pop.~III stars in the central neighborhood of a halo, at a radial distance of a few parsecs. We use an NFW halo profile~\cite{Navarro:1995iw} to describe the DM density, with parameters drawn from N-body simulations. The local DM density at this distance turns out to be $\rho_\chi=5.3\, {\rm GeV/cm^3}$, which is several orders of magnitude smaller than the densities considered in the literature cited above. We model the DM velocity distribution using the Eddington inversion method~\cite{Eddington:1916}, since it gives a sounder approximation, closer to the \emph{real} distribution at this distance to the center of a halo than a Maxwell-Boltzmann velocity distribution function~\cite{Lopes:2020dau}. 
We compare the two approaches and find that the low-velocity tail is suppressed relative to the Maxwell–Boltzmann assumption, resulting in a lower capture rate.

Aside from modeling the DM velocity distribution using the Eddington inversion method rather than a Maxwell-Boltzmann distribution, we incorporate several key improvements over previous analyses of DM capture in heavy stars. Firstly, we consider the evolution of the Pop.~III stars' composition. Secondly, we formulate a response function for hydrogen and extend the multi-scattering formalism to collisions with three different target elements. These new developments enable a more realistic calculation of DM capture in stars. 

Finally, we have found that captured DM can thermalize in the core of the first stars across a wide region of the parameter space considered, and that annihilating DM rapidly reaches capture–annihilation equilibrium, remaining dynamically irrelevant. Non-annihilating DM, on the other hand, can accumulate, thermalize, become self-gravitating, and in parts of parameter space collapse into a black hole capable of destroying the star before the end of its life. 
 All of this under the assumption of relatively low DM densities in the vicinity of the star.  
 These results highlight the critical roles of stellar evolution and internal composition in assessing the impact of heavy DM on massive stars, thereby providing a sensitive probe of heavy DM models.

This paper is organized as follows. In Section~\ref{sec:composition}, we outline the evolution and internal composition of Pop.~III stars and summarize the key structural features relevant for dark matter capture. In Section~\ref{sec:capture}, we present our treatment of DM capture, beginning with the star in the zero-age main sequence, the velocity distribution of DM in the host halo, and the resulting capture rates. We then extend the multiple-scattering formalism to include three nuclear species and evaluate capture during the late, metal-enriched stages of the star. In Section~\ref{sec:thermalization}, we discuss thermalization of captured DM in the stellar core and the subsequent annihilation dynamics. In Section~\ref{sec:collapse}, we analyze the formation of a black hole in the center of the star in the case of asymmetric DM. Finally, in Section~\ref{sec:conclusions}, we summarize our main results and conclude.

\section{Composition and evolution of Population~III stars}\label{sec:composition}

To simulate Pop.~III stars, we used the stellar evolution code Modules for Experiments in Stellar Astrophysics (\texttt{MESA})~\cite{Paxton:2011,Paxton:2013,Paxton:2015,Paxton:2017eie,Paxton:2019,Jermyn:2023}, version 23.05.1, with the same parameters as in Ref.~\cite{,Windhorst:2018wft}, which in turn relied on those given in Refs.~\cite{Farmer:2015,Fields:2016,Farmer:2016}. That is, we simulated non-rotating single stars (neither binaries nor multiple stars were considered), with masses $20,\,100$ and $1000 \Msun$, zero mass loss during their evolution, and zero metallicity $Z=0$. 
We evolved these stars from the Zero Age  Main Sequence (ZAMS), at redshift $z\sim 10$, up to the moment when helium is depleted from the star's core to a fraction of $10^{-6}$~\cite{,Windhorst:2018wft}. This is before the end of the star's life. The $20\Msun$ star is expected to undergo core collapse, triggering a supernova (SN) explosion that would leave behind a neutron star, while the others would collapse to a black hole without a SN~\cite{Klessen:2023qmc}.

\begin{figure}[t]
    \centering
    \includegraphics[width=\linewidth]{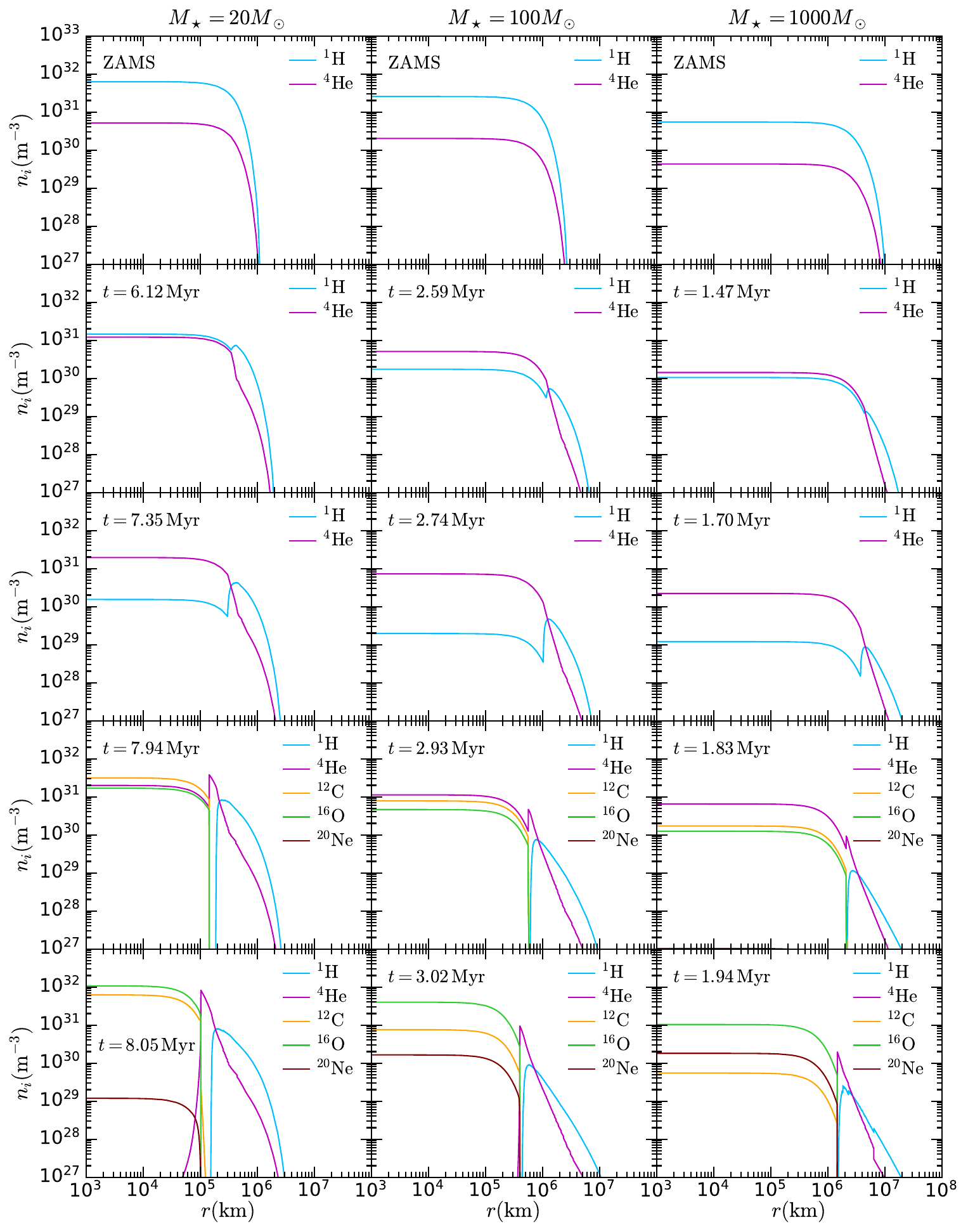}
    \caption{Evolution of the matter content of $20\Msun$,  $100\Msun$ and $1000\Msun$ Pop.~III stars from the zero age main sequence (ZAMS) until He depletion from the core, obtained using \texttt{MESA} with parameters taken from Ref.~\cite{Windhorst:2018wft}.
    }
    \label{fig:n_profiles}
\end{figure}

\begin{figure}[t]
    \centering
    \includegraphics[width=0.46\linewidth]{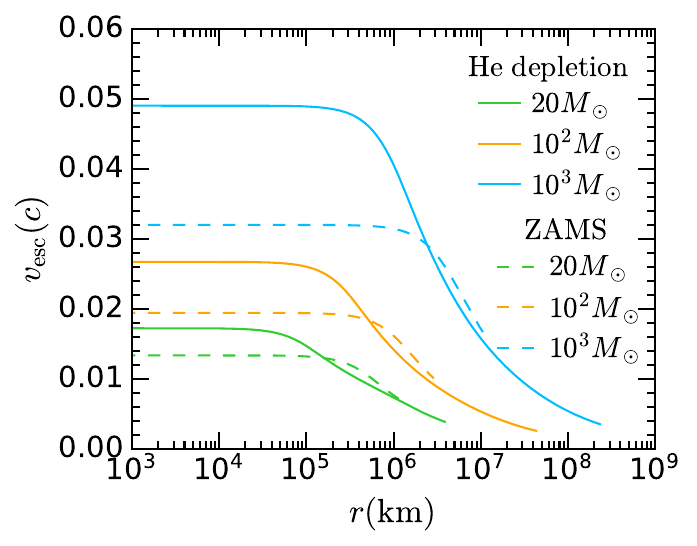}
    \includegraphics[width=0.5\linewidth]{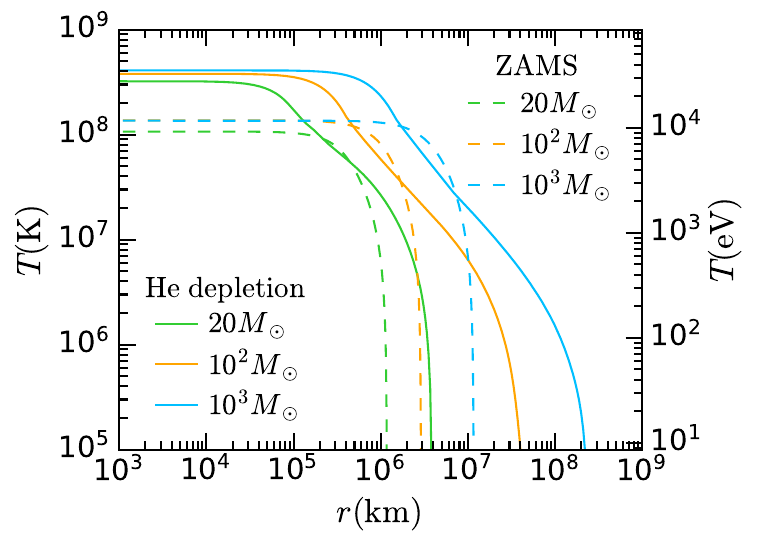}    
    \caption{Left: Escape velocity profiles at the ZAMS and He depletion stages of the Pop.~III stars presented in Fig.~\ref{fig:n_profiles}.  Right: Temperature profiles for the same stages of the same Pop.~III stars.}
    \label{fig:v_profiles}
\end{figure}

Figure~\ref{fig:n_profiles} shows the evolution of the number density $n_i$ profiles for Pop.~III stars of mass $20\Msun$, $100\Msun$, and $1000\Msun$. 
Note that there is a drastic change in the star's composition during the first Myrs in the main sequence for the $100\Msun$ and $1000\Msun$ stars. In the late stages, as metals (mainly ${}^{12}$C, ${}^{16}$O, and ${}^{20}$Ne) are formed in the interior, the star develops a metallic core and an atmosphere made of ${}^1$H and ${}^4$He. While this evolution involves practically no mass loss, the star radius, $R_\star$, changes by an order of magnitude from the main sequence to the red giant phase for the heaviest stars considered. 
This change in radius can also be seen in Fig.~\ref{fig:v_profiles}, which shows the escape velocity and temperature profiles at the zero age main sequence and at helium depletion from the core for the same stars as in Fig.~\ref{fig:n_profiles}. In the next section, we examine how these changes in composition and geometry directly affect dark matter capture. 

\section{Dark matter capture in Population~III stars}\label{sec:capture}

\subsection{Capture during the main sequence}\label{sec:early}

\begin{figure}[t]
    \centering
    \includegraphics[width=0.45\linewidth]{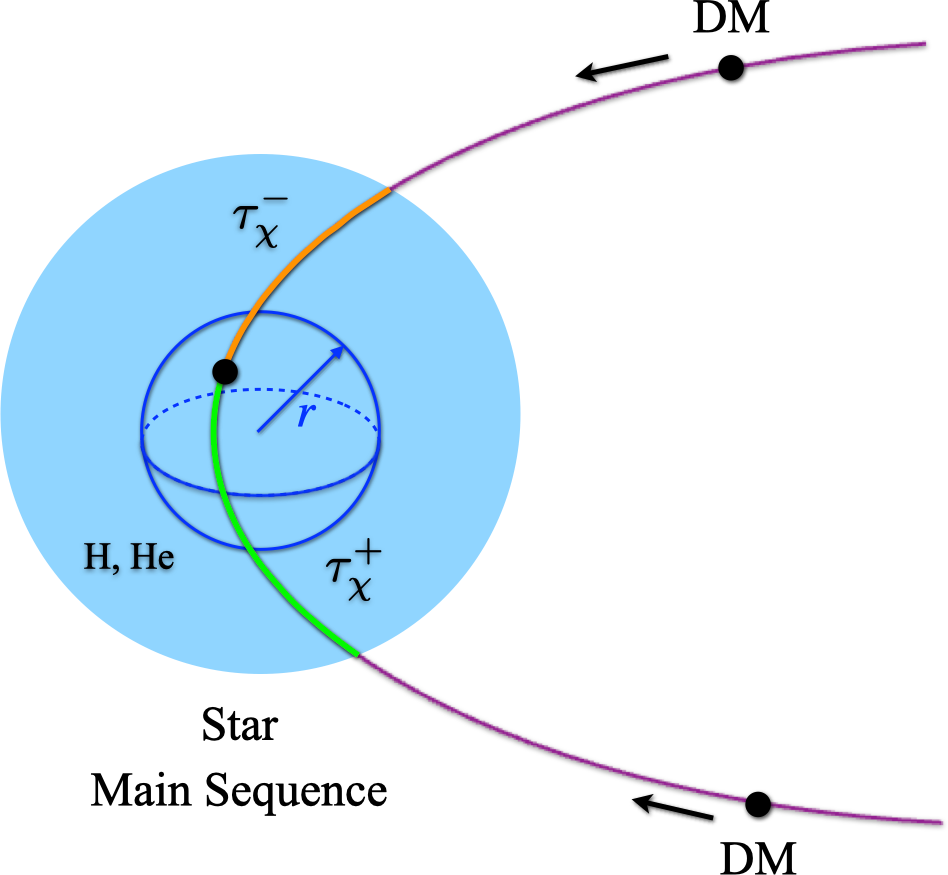}\hspace{10mm}
    \includegraphics[width=0.45\linewidth]{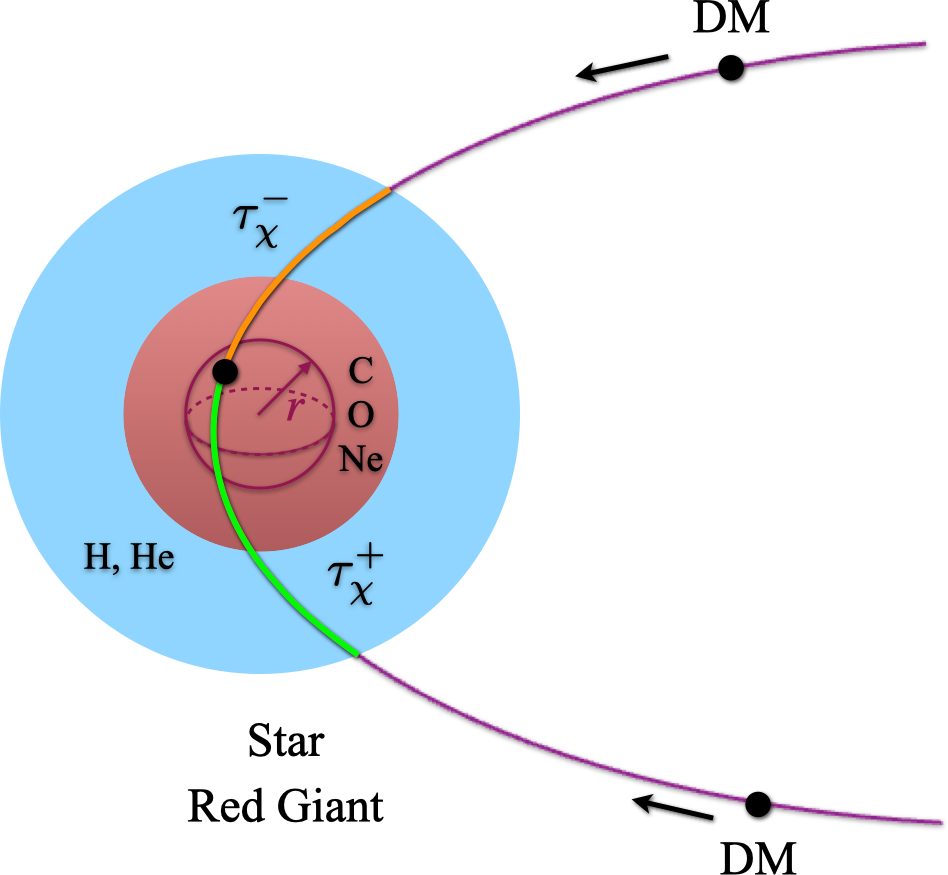}    
    \caption{Possible trajectories the DM can follow in a Pop.~III star in the main sequence (left) and red giant stage (right). The two possible paths and optical depths associated with the DM-nucleus scattering are shown. Note that in the star's late stage, the star has acquired a metal-rich core, which depending on the star mass is composed mainly of either ${}^{16}$O and ${}^{12}$C, or ${}^{16}$O and ${}^{20}$Ne, and developed a metal-free atmosphere. }
    \label{fig:trajectories}
\end{figure}

When the Pop.~III star is in the main-sequence stage, it is composed mainly of hydrogen and helium, as shown in the top row of Fig.~\ref{fig:n_profiles} and sketched in the left panel of Fig.~\ref{fig:trajectories}. At this stage, the calculation of the DM capture rate can be carried out using the formalism presented in Ref.~\cite{Bell:2024qmj} for the case of multiple scatterings involving two target species. The reason that enables us to apply that analytical approach here is that the escape velocity in a heavy Pop.~III star is comparable to that in a white dwarf~\cite{Bell:2021fye,Bhattacharjee:2025iip}, as shown in Fig.~\ref{fig:v_profiles}, and much greater than the DM velocity far from the star, as we shall see.

As the DM particle $\chi$ travels through the star, it collides with hydrogen and/or helium targets. For light DM, a single collision might be sufficient to lose enough kinetic energy and become gravitationally bound to the star after a scattering with a star constituent at a distance $r$ from the star's center. For heavy DM, multiple collisions are required. 
As sketched in Fig.~\ref{fig:trajectories}, for a DM particle with angular momentum $J$, there are two different optical depths $\tau_{\chi}^\pm(r, y)$ that are associated with a collision at a radial position $r$ from the star center, since there are two distinct paths that the particle might follow. We parametrize the angular momentum as $J\,\equiv\,y\, J_{\rm max}$, where $J_\text{max}$ is the maximum angular momentum the particle can attain.

Throughout this work, we assume that DM interacts with nucleons $N$ in the star via a scalar effective operator, which is a common interaction considered in DM direct detection analyses. The effective Lagrangian is given by
\begin{equation} \label{eq:eff_L}
    \mathcal{L}_{\rm eff} \supset \frac{c_N^S}{\Lambda^2} \overline{\chi}\chi \overline{N}N,
\end{equation} where $c_N^S$ is the corresponding nucleon hadronic matrix element and $\Lambda$ the cutoff scale of the effective theory~\cite{Bell:2021fye}.  We assume universal scalar couplings between DM and quarks, proportional to their masses, as in Higgs-portal models or Type-I Two-Higgs doublet models, so that the different proton and neutron couplings are fully captured by the coefficients $c_N^S$. Nevertheless, we present our results in terms of the DM-proton cross section. 

The probability that the DM loses at least an energy $\delta E$ via multiple scatterings with a single target species, at an optical depth $\tau_\chi$, is encoded in the response function $G(\tau_\chi(r,y),\delta)$, where $\delta\equiv \delta E/E_0$, with $E_0$ a characteristic energy associated with the target nucleus. In the case of an atom with mass number $A>1$, such as helium, the response function is given by 
\begin{equation}
\label{eq:G-function}
    G_{1,i}(\tau_\chi,\delta)\,=\, e^{-\tau_\chi-\delta}\,I_0\left[2\sqrt{\tau_\chi\,\delta}\right],
\end{equation} 
where $I_n(x)$ is the $n$-th order modified Bessel function of the first kind. Details of how to obtain this response function are found in Ref.~\cite{Bell:2024qmj} and summarized in Appendix~\ref{app:nuclear-responsefunc}. The characteristic energy, $E^i_0$, associated with the scattering of DM with a nucleus $i$, and $A>1$, is defined as
\begin{equation} \label{eq:E0s}
E^i_0\,\equiv\, \lim_{m_\chi \to\infty}\,\frac{\int_0^\infty \operatorname{d\!}E_R\,E_R \frac{\operatorname{d\!}\sigma_{i\chi}}{\operatorname{d\!}E_R}}{\int_0^\infty \operatorname{d\!}E_R\, \frac{\operatorname{d\!}\sigma_{i\chi}}{\operatorname{d\!}E_R}}\, ,
\end{equation}
where $E_R$ is the recoil energy of the target $i$ and $\operatorname{d\!}\sigma_{i\chi}/\operatorname{d\!}E_R$ is the differential scattering cross section. 

On the other hand, the case of DM multiple scatterings with hydrogen (not considered in Ref.~\cite{Bell:2024qmj}) is exceptional since its nuclear form factor is trivial. In this case, the response function is given by 
\begin{align} 
\label{eq:GH-function}
     G_{1,i}(\tau_\chi, \delta )&=  \sum_{m=0}^\infty \Theta (\delta-m)\frac{ (-1)^{m+1} e^{-\tau_\chi} \sqrt{\tau_\chi} }{\Gamma (m+1)}\left(\sqrt{\tau_\chi(\delta -m)}\right)^{m-2}\\
     &\times  \left((\delta -1) \sqrt{\delta -m}
    \, I_{m-1}\left[2 \sqrt{\tau_\chi(\delta -m)}\right]+\sqrt{\tau_\chi}
    (m-\delta ) I_{m-2}\left[2 \sqrt{\tau_\chi(\delta -m)}\right]\right)\, , \nonumber
 \end{align} 
 where $\Theta(x)$ is the Heaviside step function. For the hydrogen case, $E_0^i$ is the maximum energy transfer in a single collision. The steps taken to arrive at this response function are presented in Appendix~\ref{app:H-responsefunc}. As explained in the Appendix, since this is a series of exponentially oscillating terms of alternating sign, for the numerical calculations, we have implemented a piecewise definition of the response function, interpolating the regimes in which $\delta$ is much larger and smaller than unity, together with the regime where $\tau_\chi \sim 2\delta$, in which case the energy lost through the optical length $\tau_\chi$ is about the average single-collision energy loss. 

In the case of two target elements, the probability for DM to lose an energy $\delta E$ after colliding with the species $i$ and $j$, with optical depths $\tau_\chi^i$ and $\tau_\chi^j$, respectively, is 
\begin{equation}
\label{eq:G2-function}
    G_{2,ij}(\tau_\chi^i,\tau_\chi^j,\delta E)\,=\, \int_0^{\delta E/ E_0^j}\operatorname{d}\!z\,G_{1,i}\left(\tau^i_\chi,\,\frac{\delta E - z E^j_0}{E^i_0}\right)\left[\frac{\partial}{\partial z}\int_{0}^{\tau_\chi^j} \operatorname{d}\!\tau \,G_{1,i}(\tau_\chi,z) \right],
\end{equation}
where $G(\tau_\chi,\delta)$ is given by Eqs.~\eqref{eq:G-function} or \eqref{eq:GH-function}, depending on the corresponding target. 

Now, we can write the probability that the DM particle loses an energy $\delta E$  as the sum of the probability for multiple scatterings with only one target $i$ plus the probability of collisions with two targets $i$ and $j$. This is given by
\begin{equation} 
\label{eq:G12-function}
    G_{12,ij}(r,\delta_i)\,=\, G_{1,i}(\tau^i_\chi,\,\delta_i) e^{-\tau^j_\chi}+G_{2,ij}(\tau_\chi^i,\tau_\chi^j,\delta E). 
\end{equation} Finally, we average over the DM particle's angular momentum and the two possible trajectories, $\tau_\chi^+$ and $\tau_\chi^-$,
\begin{equation}
   \tilde{G}_{12,ij}(r,\delta_i)\,=\,\int_0^1 \operatorname{d}\!y \frac{y}{\sqrt{1-y^2} } \frac{G_{12,ij}(\tau^+_\chi\,\delta_i)+G_{12,ij}(\tau^-_\chi\,\delta_i)}{2}.
\end{equation}

The multi-target multi-scattering capture rate when the DM particle scatters mainly with the target $i$ is then given by
\begin{equation} \label{eq:capture_target_i}
C_i\,=\, \frac{\rho_\chi(r_\star)}{m_\chi}\,\int^{R_\star}_0 \operatorname{d}\!r \,4\pi r^2 n_i(r) \,v_{\rm esc}^2(r) \int^{v_{\rm esc}^{\rm halo}(r_\star)}_0 \frac{\operatorname{d}\!u_\chi}{u_\chi} f_{\chi\star}(u_\chi) \,\sigma_{i\,\chi}(r) \,\tilde{G}_{12,ij}(r,\delta_i)\,,   
\end{equation}
where, $u_\chi$ is the DM speed away from the star, $v_{\rm esc}(r)$ is the radially dependent escape velocity from the star, $f_{\chi\star}(u_\chi)$ is the velocity distribution function in the star's reference frame, $\rho_\chi(r_\star)$ is the DM density at a radial distance $r_\star$ from the halo's center,  $v_{\rm esc}^{\rm halo}(r_\star)$ is the halo escape velocity at the same distance, and $\sigma_{\chi \,i}(r)$ is the scattering cross section of DM with the target $i$, which, for the DM-target interaction we are considering, depends on the radial position in the star through the nuclear form factor.
The differential scattering cross section is given in Appendix~\ref{app:form-factors} together with the nuclear form factors taken from Ref.~\cite{Catena:2015uha} that are embedded in the differential cross section.

After calculating the rate in Eq.~\eqref{eq:capture_target_i} for each of the targets involved, we obtain the total capture rate by adding the individual contributions,
\begin{equation} \label{eq:total_capture}
C\,=\, C_i\,+\,C_j\, .
\end{equation}

Finally, it is worth mentioning that since we are focusing on the capture of DM with mass above the TeV scale, we have safely neglected the motion of the targets whose thermal energy is much lower than the DM kinetic energy (see right panel of Fig.~\ref{fig:v_profiles}), i.e.,  we are working in the zero temperature approximation. 

\subsubsection{The dark matter velocity distribution} \label{sec:vel_dist_func}

\begin{figure}[t]
	\centerline{\includegraphics[width=0.5\textwidth]{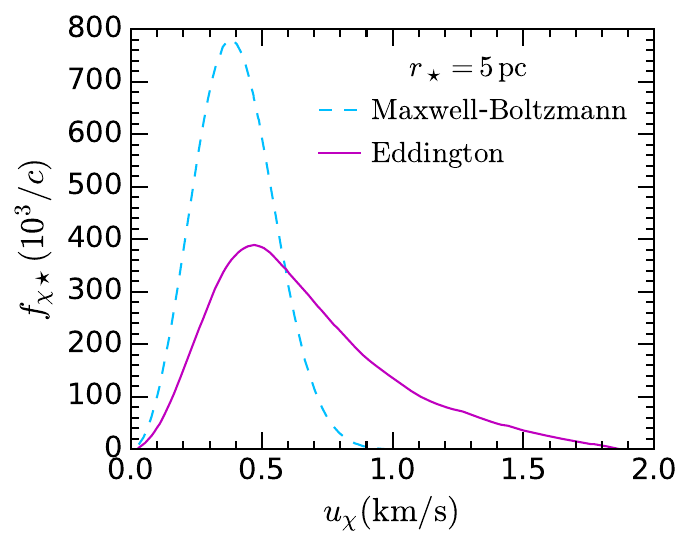}
	}
	\caption{Dark matter velocity distribution function in the star's frame,  obtained using the Eddington inversion method~\cite{Eddington:1916} for a $10^5\Msun$ halo at redshift $z=10$ with a DM density profile in Eq.~\eqref{eq:NFW_profile} (solid magenta line), and for a star at a radial distance $r_\star=5{\rm \, pc}$ from the center of the halo. For comparison, we also show a Maxwell-Boltzmann velocity distribution using $v_\star\,=\,v_d\,=\,v_{\rm circ}(r_\star=5{\rm \, pc})$ (dashed light blue line).} \label{Fig:Vel_dist}
\end{figure}

In the definition of the capture rate, Eq.~\eqref{eq:capture_target_i}, $f_{\chi\star}(u_\chi)$ is the velocity distribution function of the dark matter particles in the frame of the star. Most of the literature on DM capture in celestial bodies assumes that the DM velocity follows a Maxwell-Boltzmann (MB) distribution function,
\begin{equation} 
\label{eq:fMB}
f_{\rm MB}(u_\chi)\,=\,\frac{u_\chi}{v_d\,v_\star}\sqrt{\frac{3}{2\pi}}\left[\operatorname{Exp}\left(-\frac{3}{2v_d^2}\left(u_\chi-v_\star\right)^2 \right)-\operatorname{Exp}\left(-\frac{3}{2v_d^2}\left(u_\chi+v_\star\right)^2 \right) \right],
\end{equation} where $v_\star$ is the star's velocity and $v_d$ is the velocity dispersion. This is a reasonable assumption for calculations in the vicinity of the solar system, at approximately 8~kpc from the center of the Galaxy. However, it has been pointed out that, for much shorter distances, Eq.~\eqref{eq:fMB} leads to an overestimation of the low-velocity population~\cite{Lopes:2020dau}. Consequently, we use the Eddington inversion method~\cite{Eddington:1916} (outlined in Appendix~\ref{app:Eddington}), which makes use of the DM density profile and total gravitational potential of the halo to obtain a more realistic velocity distribution function~\cite{Lopes:2020dau}. We model the halo DM density with an NFW profile~\cite{Navarro:1995iw} and use the fit of the concentration $c(z)$ mass relation from N-body simulations as a function of redshift $z$,  given in Ref.~\cite{Dutton:2014}, to obtain the relevant NFW parameters, i.e. scale radius  $r_s$  and its corresponding density $\rho_s$, as follows
\begin{align} 
\label{eq:NFW_profile}
a(z) &= 0.520+0.385\exp{(-0.617 z^{1.21})}, \\
b(z) & = -0.101 + 0.026 z,\\
\log{c(z)} &= a(z)+b(z)\left( \frac{M_\text{halo}}{10^{12} M_\odot} \right), \\
r_s(z) &= \frac{R_\text{vir}(z)}{c(z)}, \qquad R_{200}^3 = \frac{3 M_\text{halo}}{800\pi \rho_\text{crit}(z)},  \\
\rho_s(z) & = \frac{200}{3} \rho_\text{crit}(z) \frac{c^3}{\ln(1+c)-\frac{c}{a+c}},
\end{align}
where the virial radius $R_\text{vir}$ has been taken to be $R_{200}$ (the radius for which $\rho_\chi=200\,\rho_\text{crit}$). We have taken    $M_\text{halo}=10^5 M_\odot$. This fit can be used for radial distances from the center of the halo in the range $\sim 0.1\, \text{pc} - R_\text{vir}(z)$.

In Figure~\ref{Fig:Vel_dist}, we show the resulting distribution function in the reference frame of a star near the center of a $10^5\Msun$ halo at $z=10$, namely at a distance $r_\star\sim 5\, {\rm pc}$ from the halo's center.  The DM density at this distance is $\rho_\chi\,\approx\,5.3\,{\rm GeV/cm^3}$. We also include the velocity distribution obtained using the standard Maxwell-Boltzmann expression~\eqref{eq:fMB}, for comparison. Our results indicate that, at this distance from the halo's center, the Maxwell-Boltzmann approach overestimates the low-velocity tail of the distribution, which is the most relevant regime for calculating DM capture rates.

\subsubsection{Zero age main sequence results}

\begin{figure}[t]
	\centerline{\includegraphics[width=0.65\textwidth]{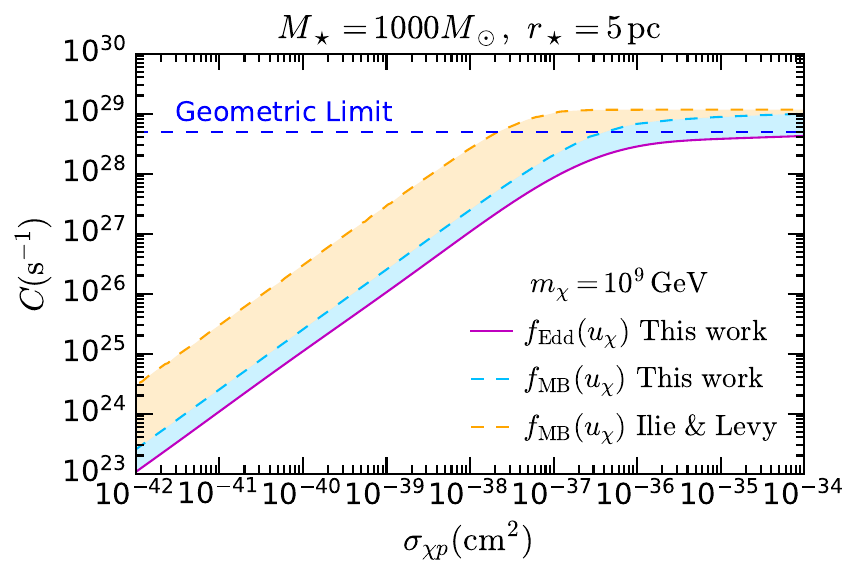}
	}
	\caption{DM capture rate in the zero age main sequence (ZAMS) stage of a Pop.~III star with $M_\star=1000 \Msun$ at 5 pc from the halo center. The solid magenta line represents the total capture rate calculated using Eq.~\eqref{eq:total_capture} and the DM velocity distribution obtained using 
    the Eddington inversion method (see Fig.~\ref{Fig:Vel_dist}). The dashed light-blue line represents the total capture obtained using the same procedure but with a Maxwell-Boltzmann velocity distribution, Eq.~\eqref{eq:fMB}. Finally, the dashed orange line shows the result obtained using the formalism in Ref.~\cite{Ilie:2021iyh}.} \label{Fig:DM_Capture_comparison}
\end{figure}

We have computed the total capture rate for the initial stage of the Pop.~III star. We have used the benchmark values $v_\star\sim v_d \,=\,0.3\,{\rm km/s}$, which correspond to the circular velocity at $r_\star=5\,{\rm pc}$ in the gravitational potential generated by the halo given in Eq.~\eqref{eq:NFW_profile}. Fig.~\ref{Fig:DM_Capture_comparison} shows the total capture rate in the ZAMS stage of the $1000\Msun$ Pop.~III star, for $m_\chi=10^9\,{\rm GeV}$. The figure shows that, for large values of the DM-proton cross section, the capture rate reaches a saturation value that corresponds to the so-called geometric capture rate, meaning that all DM particles entering the star are captured. The geometric capture rate is defined as
\begin{equation}\label{eq:C_geom}
    C_{\rm geom}\,=\,\frac{\pi R_\star^2\, \rho_\chi(r_\star)}{m_\chi} \int_0^{v_{\rm esc}^{\rm halo}(r_\star)} \operatorname{d}\!u_\chi \frac{v_{\rm esc}^2(R_\star)+u_\chi^2}{u_\chi} \, f_{\chi\star}(u_\chi)\, ,
\end{equation} where $f_{\chi\star}(u_\chi)$ is calculated using the Eddington inversion method, as discussed in section \ref{sec:vel_dist_func}.

For comparison, Fig.~\ref{Fig:DM_Capture_comparison} shows the capture rate calculated using a Maxwell-Boltzmann velocity distribution (dashed light-blue line), Eq.~\eqref{eq:fMB}. The figure also shows the results obtained using the formalism presented in Ref.~\cite{Ilie:2021iyh} (dashed orange line), which also assumes an MB speed distribution. While all results exhibit similar behavior, it is noticeable that the transition from the optically thin regime to the saturation of the geometric limit is less sharp in our results compared to the method in Ref.~\cite {Ilie:2021iyh}, and occurs at an order of magnitude larger DM-proton cross section. As a reminder, we consider the radial dependence of the stellar composition and the escape velocity, rather than assuming constant values. Assumptions in previous work lead to an overestimation of the capture rate by approximately one order of magnitude for scattering cross sections below the saturation limit. Finally, we notice that there is an additional overestimation from using the Maxwell-Boltzmann velocity distribution  (compare the dashed light blue with the solid magenta line) since the low-velocity tail is more prominent in that distribution, as seen in Fig.~\ref{Fig:Vel_dist}.

\begin{figure}[t]
\centering
    \includegraphics[width=\linewidth]{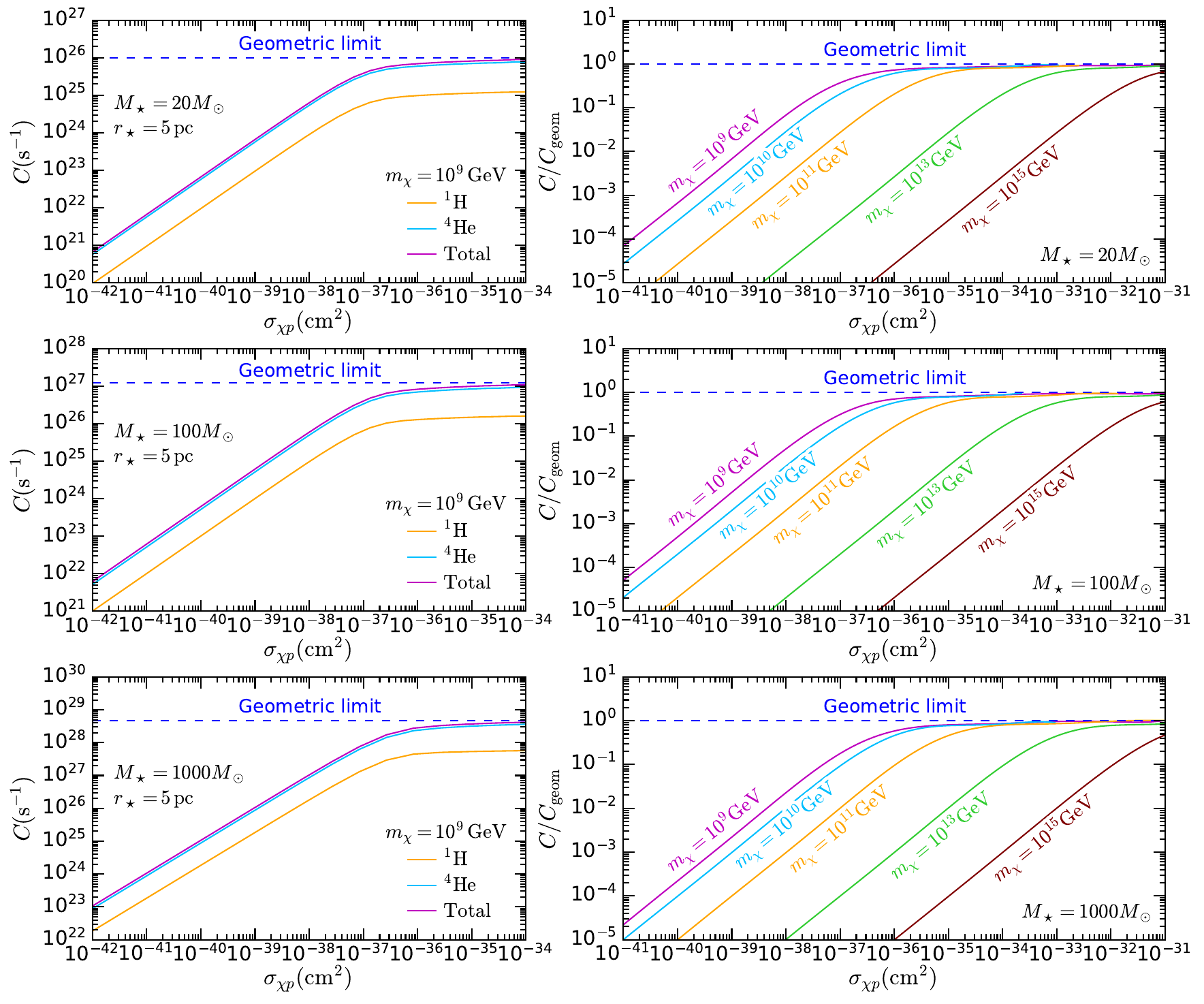}
	\caption{{\it Left:} Capture rate, as a function of the DM-proton scattering cross section for  $m_\chi\,=\,10^{9}$, in the early stage (ZAMS) of the three stars with masses $M_\star\,=\,20,\,\,100,\,\,$ and $1000\,M_\odot$ at a distance $r_\star=5$~pc from the halo center. The contributions of H and He to the total capture rate are shown in different colors. 
     {\it Right:} Ratio of the capture rate to the geometric limit in the late stage of the same stars for various values of $m_\chi$. } 
    \label{Fig:DM_CapOverCg_early_3stars}
\end{figure}

The contributions of DM scattering with H and He to the total capture rate for the three Pop.~III stars considered here, and DM of mass $10^9\,$~GeV are shown in the left panels of Fig.~\ref{Fig:DM_CapOverCg_early_3stars}.   As expected, the capture rate is dominated by the helium contribution despite its underabundance and a nontrivial form factor that suppresses the corresponding cross section. This is due to coherent scattering, with the DM-nucleus cross section being enhanced by a factor $\sim A^4$, where $A$ is the atomic number. We also note the increase in the capture rate with the stellar mass from $M_\star=20\Msun$ to $1000\Msun$. 
In the right panels of the figure, we show the ratio of the total capture rate to the corresponding geometric limit. We notice that the ratio $C/C_{\rm geom}$ is the same for all masses $m_\chi\,\leq\,10^9\,{\rm GeV}$. In that regime, single-scattering capture dominates and the rate scales as $C \sim m_\chi^{-1}$. On the other hand, for $m_\chi\,>\,10^9\,{\rm GeV}$, multi-scattering effects are necessary and the capture rate scales as $C \sim m_\chi^{-2}$. 
In Appendix~\ref{sec:scaling}, we expand on the details about this scaling. Finally, it is notable that capture due to collisions with hydrogen is significantly less dominant than that with helium.

\subsection{Capture during the late red giant stage: Scattering with three different elements}

The results from our \texttt{MESA} simulations show that, after a few Myrs., the Pop.~III stars have developed a core with a large abundance of ${}^{12}$O and other metals (see Fig.~\ref{fig:n_profiles}), while keeping an atmosphere with ${}^4$He and ${}^1$H, as sketched in the right panel of Fig.~\ref{fig:trajectories}. The presence of more than two target species in non-negligible amounts requires us to extend the formulation of the response function to include the case when the DM particle is captured after collisions with three different elements, $i,\,j\,$ and $k$. To this end, we follow the same rationale as Ref.~\cite{Bell:2024qmj} to extend the formalism discussed in Section~\ref{sec:early} to the case of three distinct target species. When the DM particle interacts more predominantly with the target $i$, but also collides with the targets $j$ and $k$, the probability for the DM particle to be captured at a depth between $\tau^i_\chi$ and $\tau^i_\chi+\operatorname{d\!}\tau^i_\chi$ is then given by
\begin{align} 
\label{eq:G3_function}
    G_{3,ijk}(\delta E)\,=\, \int_0^{\delta E/E^j_0}\operatorname{d}\!z_1 &\int_0^{\frac{\delta E-z_1E_0^j}{E^k_0}}\operatorname{d}\!z_2  G_{1,i}\left(\tau^i_\chi , \frac{\delta E - z_1 E_0^j-z_2E_0^k}{E_0^i}\right) \nonumber\\
    &
    \times \left[\frac{\partial}{\partial z_1}\int_{0}^{\tau_\chi^j} \operatorname{d}\!\tau_\chi \,G_{1,j}(\tau_\chi,z_1) \right] \left[\frac{\partial}{\partial z_2}\int_{0}^{\tau_\chi^k} \operatorname{d}\!\tau_\chi \,G_{1,k}(\tau_\chi,z_2) \right], 
\end{align}
where $G_{1,i}(\tau_\chi,z)$ is given in Eq.~\eqref{eq:G-function}, and $E_0^i$, $E_0^j$, and $E_0^k$ are characteristic energies associated with the targets, defined in Eq.~\eqref{eq:E0s}.

The DM particle may be captured after collisions with only one type of element $i$, or two $ij$, $ik$, or three $ijk$. Remembering that the probability that the DM particle does not interact with target $i$ is given by $e^{-\tau^i_\chi}$, the total probability for the particle to lose an energy $\delta E$ is then given by 
\begin{eqnarray} \label{eq:G_tot_3targets}
G_{\rm tot}(\delta_i)&=& G_{1,i}(\tau_\chi^i,\delta_i)e^{-\tau^j_\chi-\tau^k_\chi}+G_{2,ij}(\delta E)e^{-\tau^k_\chi}+G_{2,ik}(\delta E)e^{-\tau^j_\chi}+G_{3,ijk}(\delta E),
\end{eqnarray}
 where the $G_{1,i}(\tau_\chi^i,\delta_i)$, $G_{2,ij}(\delta E)$, and $G_{3,ijk}(\delta E)$ are given in Eqs.~\eqref{eq:G-function}, \eqref{eq:G2-function}, and \eqref{eq:G3_function}, respectively.

\begin{figure}[t]
    \centering
    \includegraphics[width=0.75\linewidth]{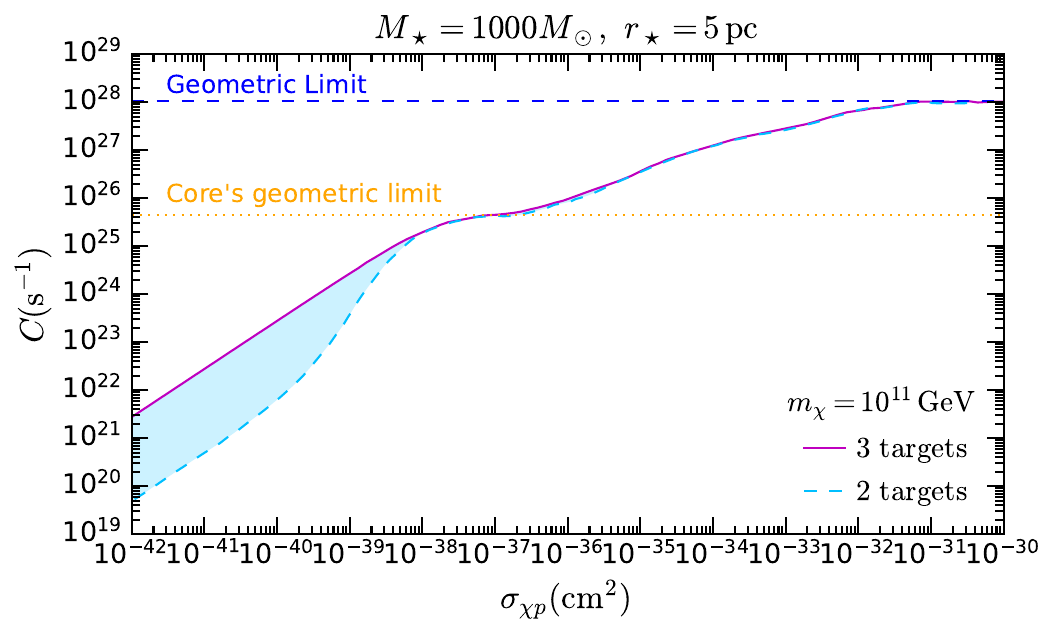}    
    \caption{Comparison between using three and two target elements to calculate the capture rate of heavy DM during the final stage of the star. The solid line shows the results that include collisions with three target species. The dashed line corresponds to the capture rate considering only collisions with two target elements. There is a clear deficit in the latter with respect to the former. This difference is significant for $m_\chi \gtrsim 10^9 \,{\rm GeV}$ and becomes more pronounced for heavier DM particles. }
    \label{fig:comparison2-3-targets}
\end{figure}

Figure~\ref{fig:comparison2-3-targets} highlights the importance of including scattering with three different target species in a $1000\Msun$ Pop.~III star (O$+$Ne$+$He or O$+$Ne$+$H). 
Several features are worth noting in the figure. First, the presence of a distinct core can be immediately recognized as the capture rate approaches a first stage of saturation at $\sigma_{\chi p}\sim10^{-37}{\rm cm}^2$, which is given by the geometric capture limit due to a sphere of the core's size. For larger scattering cross sections, any DM particle going through the core will be captured, and the total capture rate keeps increasing until reaching the star's geometric limit. At that point, any DM particle entering the atmosphere will also be captured\footnote{A similar effect was found in the capture of WIMP-like DM in low mass red giants in globular clusters~\cite{Hong:2024ozz}.}. 
The dashed line represents the capture rate calculated assuming only two target elements (O$+$Ne in the core, and He$+$H in the atmosphere). We observe a significant deficit in the rate at small DM-proton cross sections when we limit the calculation to this case, see the shaded region. We found that this effect is present for $m_\chi>10^9\,{\rm GeV}$, which marks the transition into the regime where multiple scatterings cannot be ignored, and becomes more pronounced for heavier DM masses. The difference can be understood as follows. For heavier DM masses and small cross sections, the particle can cross the atmosphere on its way in, losing only a small amount of kinetic energy, and continue through the core, interacting with the metals. However, due to the smallness of $\sigma_{\chi p}$, the core is not optically thick, and the particle may not be trapped in the core but exit back to the atmosphere, where it is trapped after depositing the rest of its kinetic energy. Therefore, it is necessary to use the three-target response function. Recall that the contribution of scattering with hydrogen is negligible with respect to that of helium (see Fig.~\ref{Fig:DM_CapOverCg_early_3stars}). As we move towards larger cross-sections, the interaction is enough for the particle to be trapped in the core, and there is no difference between the two- and three-target calculations. While the figure only shows the case of $M_\star=1000\Msun$, the same effect is observed in the lighter stars of $M_\star=20\Msun$ and $100\Msun$.

\begin{figure}[t]
    \centering
    \includegraphics[width=\textwidth]{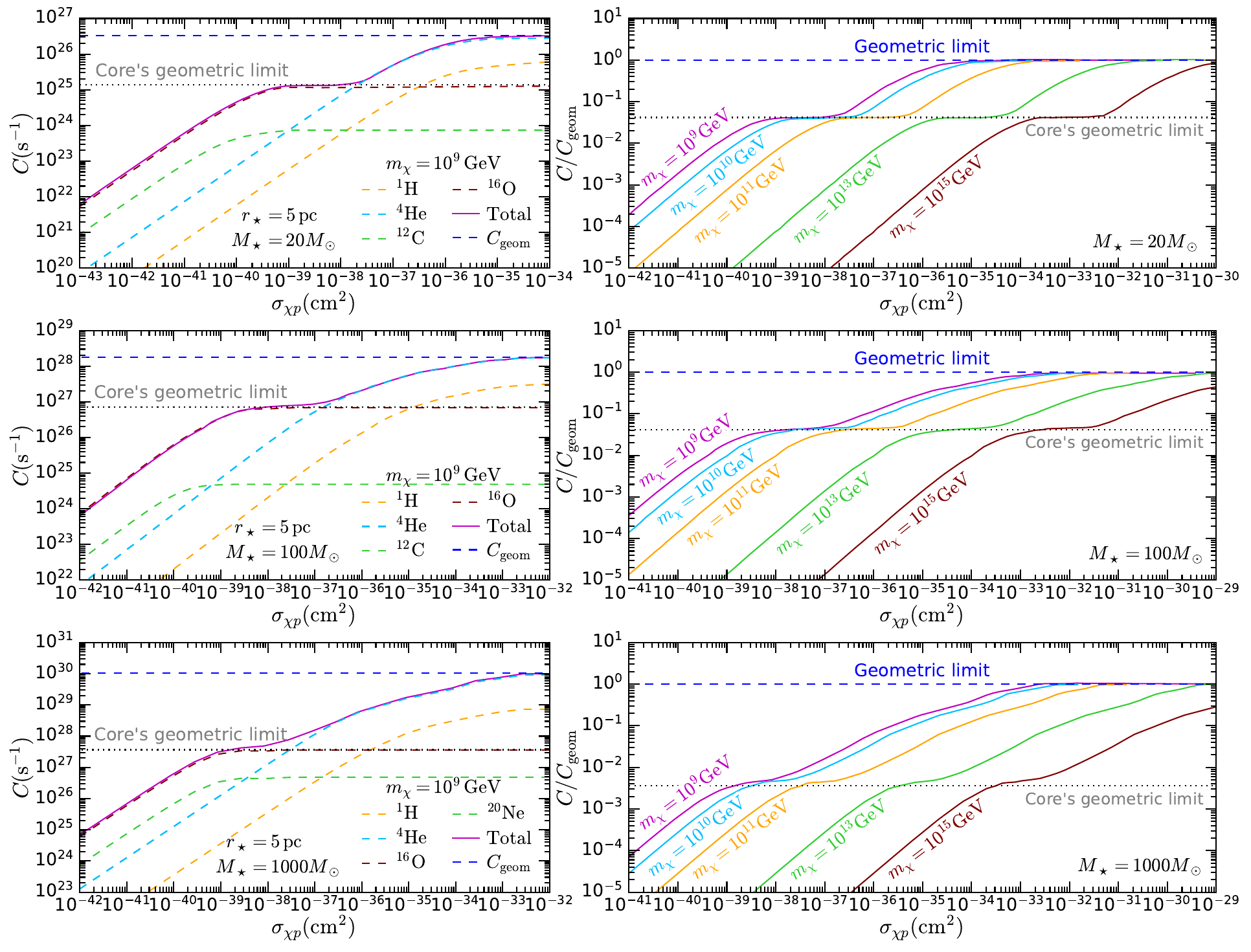}
    \caption{{\it Left:} Capture rate, as a function of the DM-proton scattering cross section, for  $m_\chi\,=\,10^{9} \,{\rm GeV}$ in the late stage of the three stars (He depletion from the core), with $M_\star\,=\,20,\,\,100,\,\,$ and $1000\,M_\odot$  at a distance $r_\star=5$~pc from the halo center. In all cases, a metallic core is present. As the DM-proton scattering cross section increases, the capture rate saturates the geometric limit of the star's core, then saturates the star's geometric limit. The contributions of each element present in the star to the total capture rate are shown in different colors. 
 {\it Right:} Ratio of the capture rate to the geometric limit in the late stage of the same stars for several DM masses.}
    \label{Fig:DM_Capture_latevs_sigma}
\end{figure}

Further insight into the different target elements in the core and in the envelope of massive Pop.~III stars, contributing to the total capture rate, can be found in the left panels of Fig.~\ref{Fig:DM_Capture_latevs_sigma}, where we show the capture rate for a DM particle with mass $m_\chi\,=\,10^{9}\,{\rm GeV}$. In all cases, we consider trajectories involving collisions with up to three different elements, using Eq.~\eqref{eq:G_tot_3targets}. For the three-target contribution, we use the three most abundant elements. The Pop.~III stars of mass $20\,M_\odot$ and $100\,M_\odot$ have carbon and oxygen as the elements in the core responsible for the capture of DM at small cross sections, whereas for the $1000\Msun$ star, the primary targets in the core contributing are oxygen and neon. In the atmosphere, helium is the most abundant element and, as in the early stages of the star, the contribution of hydrogen to capture is very small. The right panels of Fig.~\ref{Fig:DM_Capture_latevs_sigma} show $C/C_{\rm geom}$ as a function of the $\sigma_{\chi p}$ for different DM masses. Similarly to the ZAMS case, this ratio coincides for masses below $m_{\chi}\sim 10^9\,{\rm GeV}$, while scaling with the DM mass for heavier values. 
Note that more DM is captured in the late stage of the Pop.~III stars, compare left panels of Figs.~\ref{Fig:DM_CapOverCg_early_3stars} and \ref{Fig:DM_Capture_latevs_sigma}, this is not only due to the increase in the star radius (see Figs.~\ref{fig:n_profiles} and \ref{fig:v_profiles}), but also to the metallic core made of heavier elements. The latter can be observed by comparing the right panels of the same figures, where we note that for cross sections smaller than that required to reach the ``core's geometric limit'',  the capture rate is a higher fraction of the geometric limit in the late stage.

\section{Dark matter thermalization and annihilation} \label{sec:thermalization}
As captured DM particles interact with ordinary matter in the star, losing energy in the process, they thermalize in the innermost region of the stellar core. Near the center of the star, density and temperature are approximately uniform, as shown in Figs.~\ref{fig:n_profiles} and \ref{fig:v_profiles}. 

On the other hand, massive stars evolve rapidly, particularly in their late stages, as shown in Fig.~\ref{fig:n_profiles}. We then incorporate the time dependence of the temperature and star composition in the estimation of the thermalization time, using the approach in Ref.~\cite{Bell:2024qmj}, by solving the differential equation
\begin{equation} \label{Eq:t_therm_diffEq}
    \frac{\operatorname{d}\!x}{\operatorname{d}\! t}\,=\,E(x,t),
\end{equation} 
where $x$ is the ratio of the DM kinetic energy to the temperature in the central region of the star $T_c(t)$, and $E(x,t)$ is the transferred energy as the captured DM interacts with the star's constituents. This is given approximately by
\begin{equation}
    E(x,t)\,\sim\, \sum_{{\rm target\,}i} n^c_{i}(t) \sigma_{\chi\,i} \,\sqrt{\frac{3T_c(t)}{m_{i}}}\,\sqrt{\frac{x\,m_i}{m_{\chi}}}\left(\frac{2\,x\,m_i}{m_\chi} +\frac{\Gamma(3/2)}{\sqrt{\pi}}\right),
\end{equation} 
where $n^c_{i}(t)$ is the number density of the element $i$ at the center of the star. Notice that, unlike in previous work, including Ref.~\cite{Bell:2024qmj}, this is a time-dependent function because we account for the star's evolution.

\begin{figure}[t]
    \centering
\includegraphics[width=0.495\textwidth]{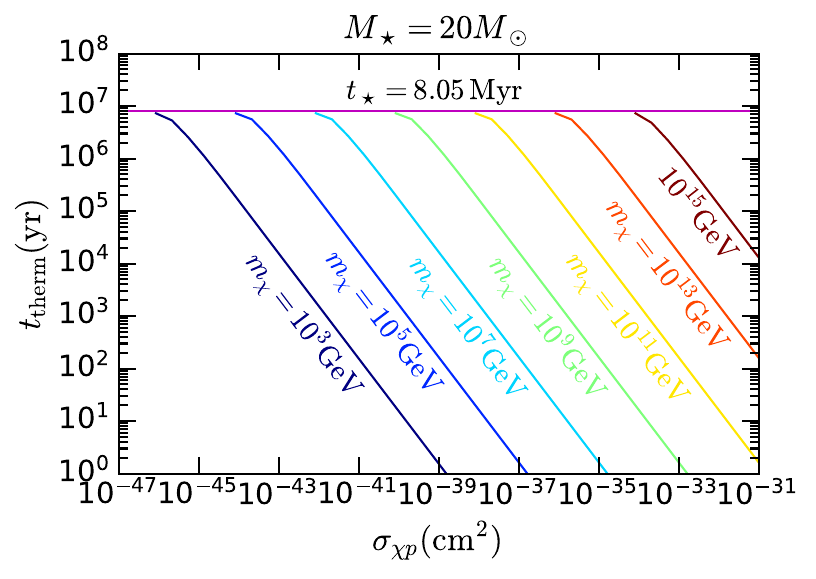}
\includegraphics[width=0.495\textwidth]{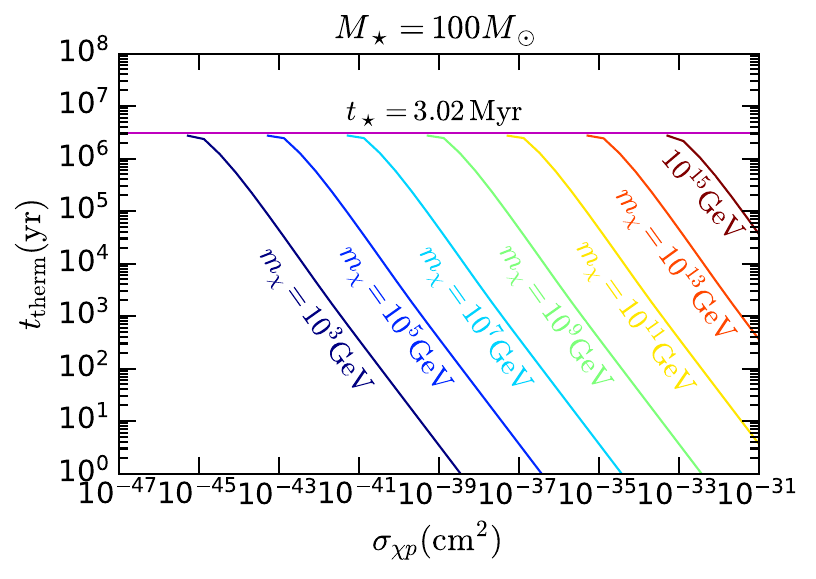}
\includegraphics[width=0.495\textwidth]{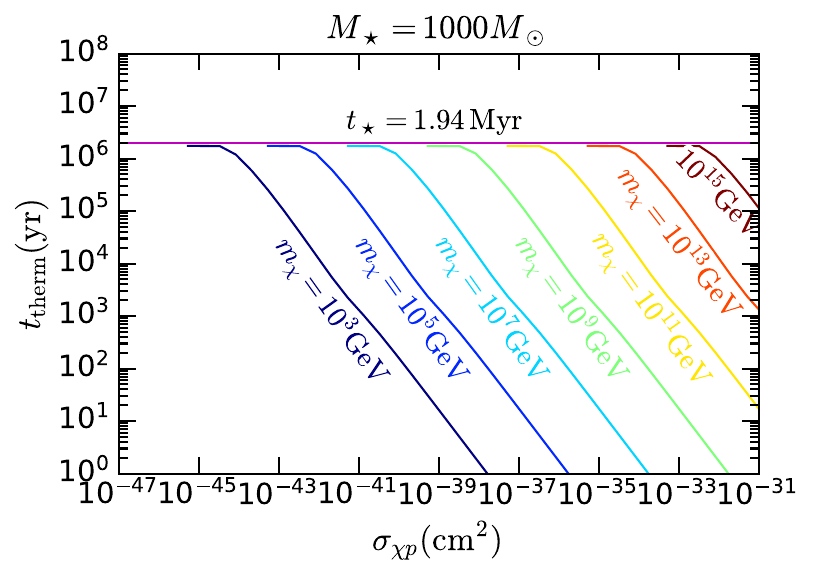}
    \caption{Thermalization time for different DM masses, increasing by two orders of magnitude from $m_\chi=10^3\,{\rm GeV}$ up to $m_\chi=10^{15}\,{\rm GeV}$. The magenta line depicts the age of the star $t_\star$ when He is depleted from its core, showing that, for some values of the DM-proton scattering cross section and the DM mass, the DM particles do not reach thermalization during the star's lifetime.}
    \label{Fig:therm_time}
\end{figure}

We can solve Eq.~\eqref{Eq:t_therm_diffEq} numerically, assuming $x\gg 1$ as an initial boundary condition, which yields the results shown in Fig.~\ref{Fig:therm_time}. Note that the previous assumption holds for DM of mass above the TeV scale. 
The figure shows the thermalization time as a function of the DM-proton scattering cross section for different values of $ m_\chi$. In general, DM with mass above the TeV scale reaches thermalization during the lifetime of the star across a wide range of cross sections, except for super-heavy DM,  $m_\chi \gtrsim10^{15}$~GeV, which would require either a time longer than the star's age or huge values of  DM-proton scattering cross section for which the DM-nucleus cross section may not be enhanced by coherent scattering~\cite{Digman:2019wdm}.

The thermalized DM occupies a sphere of radius
\begin{equation}
r_{\chi}(t)  = \sqrt{\frac{3 T_c(t)}{2\pi G m_\chi \rho_c(t)}}\, ,  
\end{equation} which is well within the region where $T_c(t)$ does not vary with $r$. The radius of the isothermal sphere remains approximately constant as the star evolves, decreasing slightly only in the last few fractions of Myr of the final stage of the star. The isothermal distribution of the DM in this sphere is given by the number density 
\begin{equation}
   n_\chi(r,t)  \simeq \frac{N_\chi(t)}{\pi^{3/2} r_\chi^3(t)} \exp{[-r^2/r_\chi^2(t)]}\, ,   
   \label{eq:niso}
\end{equation} where $N_\chi(t)$ is the number of DM particles accumulated at a given time $t$.

The thermalized DM can annihilate at a rate
\begin{equation}
 \Gamma_\text{ann}(t) =  \frac{1}{2}A(t) N_\chi^2(t),   \qquad \text{with}\,\,A\,\equiv\, \frac{\langle \sigma_{\chi\chi} v_\chi\rangle}{N_\chi^2(t)} \int \operatorname{d}^3\!r \,n^2_\chi(r,t),
\end{equation} where $\langle \sigma_{\rm \chi\chi} v_\chi\rangle$ is the thermally-averaged DM annihilation cross section. Using the  expression for $\Gamma_\text{ann}$ given in Ref.~\cite{Garani:2017jcj}, and 
 $\langle\sigma_{\chi\chi}v_{\chi}\rangle=a+bv_{\chi}^2$, a general expression for the annihilation rate reads~\cite{Bhattacharjee:2025iip}
\begin{equation}
A(t) = \frac{1}{(2\pi)^{3/2} r_{\chi}^3(t)} \left[ a + \frac{2b(3\pi-8) T_c(t)}{\pi \, m_\chi} \right] \, .
\label{eqn:Cann}
\end{equation}
Note that the annihilation cross section for the DM-nucleon interaction we are considering is $p$-wave suppressed~\cite{Zheng:2010js}. 

\begin{figure}[t]
    \centering
\includegraphics[width=0.60\textwidth]{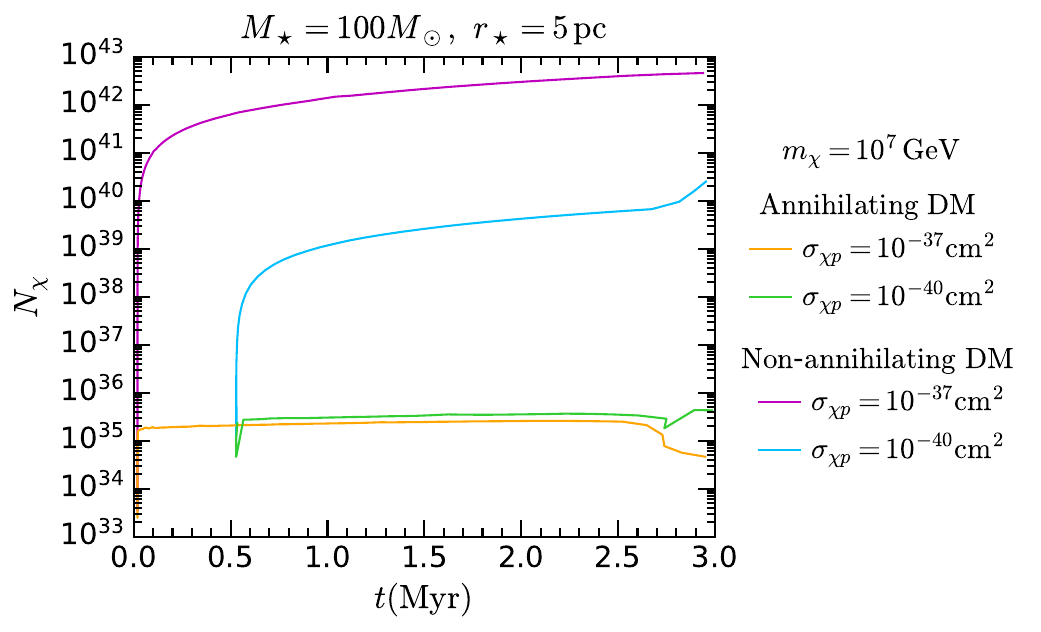}
    \caption{Number of DM particles in a $100\Msun$ Pop.~III star  at a distance $r_\star=5$~pc from the halo center as a function of time for  $m_\chi=10^7\,{\rm GeV}$, two different values of the DM-proton scattering cross section, as well as annihilating and non-annihilating DM.}
    \label{Fig:N-evolution}
\end{figure}

The number of particles in the isothermal sphere will evolve depending on the capture and annihilation rates, as prescribed by
\begin{equation}
    \frac{\operatorname{d\!}N_\chi}{\operatorname{d\!}t} = C(t) - A(t) N_\chi^2.
    \label{eq:N-t_evolution}
\end{equation} 
Note that we have omitted the evaporation term in Eq.~\ref{eq:N-t_evolution}, this is when captured DM gains energy in further scatterings and escape the star. This process that also depletes the number of accumulated DM particles is not relevant for heavy DM. 
Equation~\eqref{eq:N-t_evolution} is solved numerically, including the time dependence of the capture rate and $A$. Examples of the solutions are depicted in Fig.~\ref{Fig:N-evolution} (orange and green lines). While the capture and annihilation rates vary with time, as their rate of variation is sufficiently slow, they still reach an equilibrium, $\Gamma_\text{ann}(t) \,\simeq\, C(t)/2$ quite soon after the ZAMS stage. However, the time dependence of both processes is reflected in the fact that $N(t)$ does not reach a constant value, but evolves with the star. The annihilation of DM prevents the accumulated DM population from growing to a point that significantly affects the stellar structure.
Furthermore, the energy transferred from DM capture and annihilation $\sim C(t) m_\chi$  to a Pop.~III star of the masses considered here at a distance of 5 pc from the halo center is also not enough to alter the star's evolution, since it is well below the star's nuclear luminosity.

Next, we consider the case of non-annihilating DM, magenta and light blue lines in Fig.~\ref{Fig:N-evolution}. In the absence of annihilation, the number of DM particles accumulated in the center of the star, obtained by dropping the last term in Eq.~\eqref{eq:N-t_evolution}, is several orders of magnitude larger than in the annihilating DM case. In the next section, we consider the consequences of this scenario.

\section{Dark matter self-gravitation and collapse} \label{sec:collapse}

As the amount of DM in the star grows, the potential per DM particle, considering both the star and the DM contributions,  and  assuming a Maxwell-Boltzmann distribution for the thermalized DM, is given by~\cite{Steigerwald:2022pjo,Bell:2024qmj}
\begin{equation}
U =  -2 \pi G \rho_c(t) m_\chi r_\chi^2(t)- \frac{G m_\chi^2 N_\chi(t)}{\sqrt{2\pi} r_\chi(t)}.
\label{eq:UDM}
\end{equation}  
The DM population becomes self-gravitating when the total mass in a region of radius $r_\chi$ exceeds the mass of baryonic matter contained in the same region~\cite{Goldman:1989nd}. This occurs when the number of DM particles is
\begin{equation}
    N_\chi(t) \ge \frac{4\sqrt{2}\pi^{3/2}r_\chi^3(t)\rho_c(t)}{3\sqrt{3} m_\chi} \equiv N_\mathrm{self}. \label{eq:Nself}
\end{equation} 
Despite the large number of accumulated DM particles (see Fig.~\ref{Fig:N-evolution}), this condition is achieved only in the non-annihilating DM case. 

Once this condition is reached, DM self-gravity dominates the potential, and as more DM particles are captured and thermalized, the isothermal sphere may become unstable against gravitational collapse, which occurs when gravitational energy exceeds kinetic energy. In the case of fermionic DM, this condition is written in terms of the Fermi energy $E_{F,\chi}$,
\begin{equation}
    \frac{G m_\chi^2 N_\chi(t)}{\sqrt{2\pi} r_\chi} > E_{F,\chi},
\end{equation} where
\begin{equation}
    E_{F,\chi} = \sqrt{m_\chi^2+p_{F,\chi}^2}-m_\chi, \qquad
    p_{F,\chi} = (3\pi^2n_\chi)^{1/3}. 
\end{equation} 
This condition is similar to  the Chandrasekhar limit~\cite{Chandrasekhar1931}, which in this case leads to
\begin{equation}\label{eq:nch}
N_\chi(t)> (2)^{3/4} \pi \sqrt{3} \left(\frac{M_\text{Pl}}{m_\chi}\right)^3 = N_\text{Ch}. 
\end{equation}

After formation, the black hole (BH) at the center of the star begins accreting stellar matter. Assuming spherical accretion, the rate at which this happens is determined by 
\begin{eqnarray}\label{eq:accretion}
     \frac{\operatorname{d\!}\MBH}{\operatorname{d\!}t} \,= \,4 \pi \rho(r) v(r) r^2, \label{eq:continuity} \\
     v(r)\frac{\operatorname{d\!}v}{\operatorname{d\!}r}+\frac{1}{\rho(r)}\frac{\operatorname{d\!} P(r)}{\operatorname{d\!}r}+\frac{G\MBH}{r^2}\,=\, 0,
\end{eqnarray} where the first equation comes from the continuity equation and the second one is the Euler equation in fluid dynamics. Defining the speed of sound as $c_s^2\equiv \operatorname{d\!}P/\operatorname{d\!}\rho$, it can be shown that the unique solution to Eq.~\eqref{eq:accretion} requires the existence of a ``sonic radius'', $r_s$, at which $v(r_s)\,=\,c_s$~\cite{frank2002accretion}. Evaluating Eq.~\eqref{eq:continuity} at the sonic point, we obtain
\begin{equation}
    \frac{\operatorname{d\!}\MBH}{\operatorname{d\!}t} \,= \,4 \pi \rho(r_s) c_s r_s^2.
    \label{eq:BHaccrate}
\end{equation}

\begin{figure}[t]
    \centering
\includegraphics[width=0.495\textwidth]{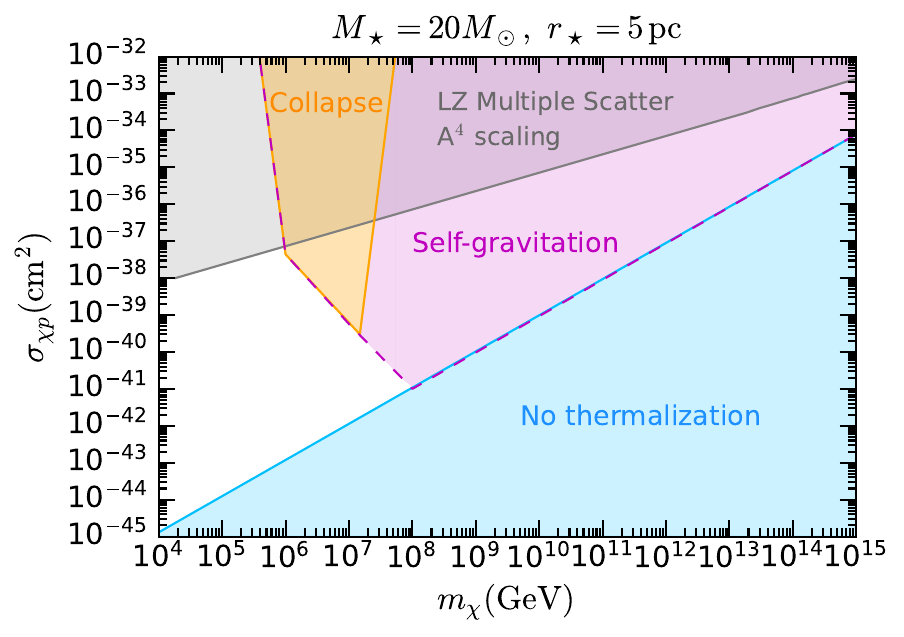}
\includegraphics[width=0.495\textwidth]{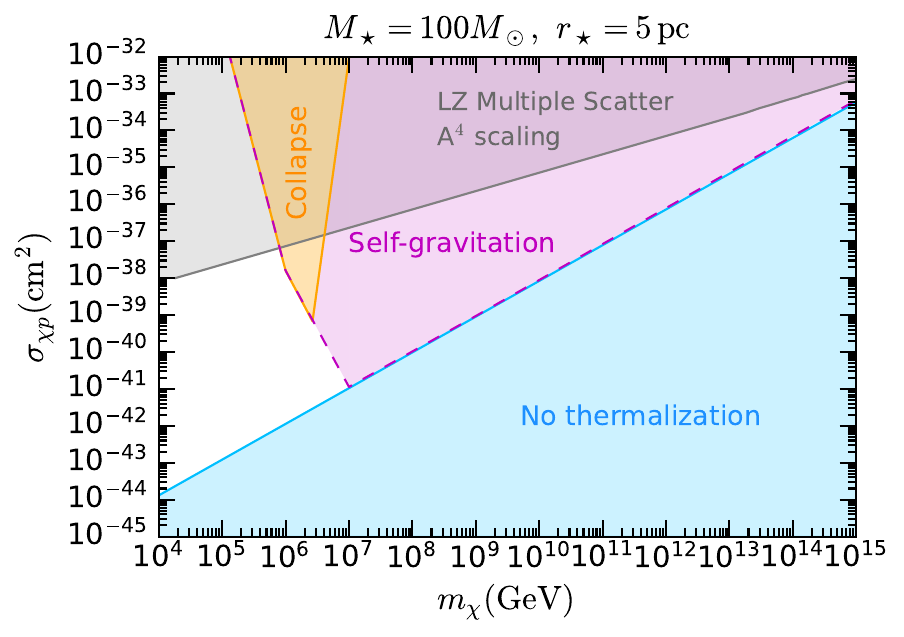}
\includegraphics[width=0.495\textwidth]{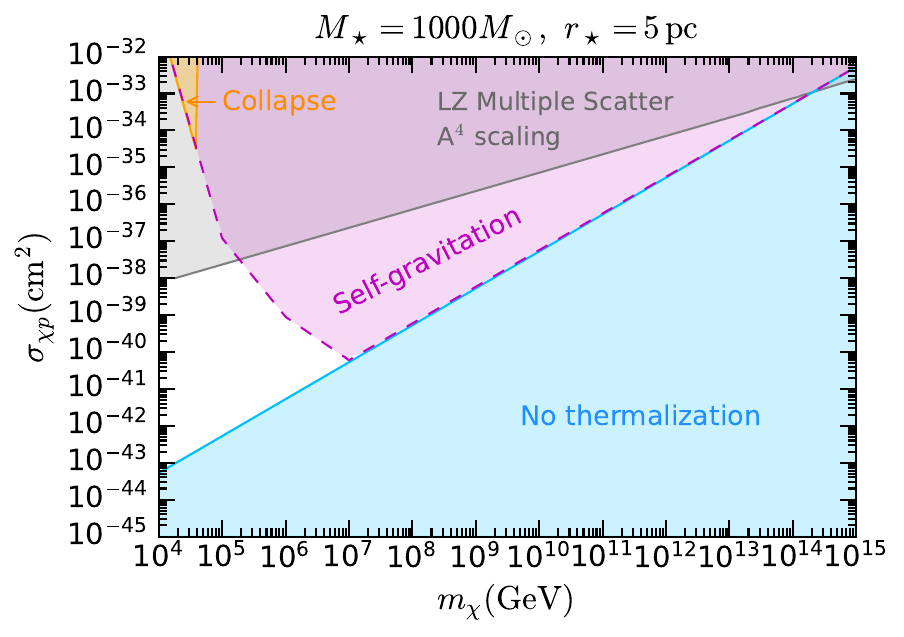}
    \caption{Regions of the parameter space where DM does not thermalize in the $m_\chi-\sigma_{\chi p}$ plane (shaded light blue area) for DM accumulated in Pop.~III stars of masses $20\Msun$ (top left panel), $100\Msun$ (top right panel), and $1000\Msun$ (bottom panel) at a distance $r_\star=5$~pc from the halo center. For the case of non-annihilating DM, we also show the regions where DM achieves self-gravitation (shaded in magenta) and those where DM gravitational collapse leads to BH formation and the star's destruction before the star reaches the end of the red giant phase. For comparison, we also show the leading bounds from direct detection experiments on heavy DM~\cite{LZ:2024psa}.}
    \label{Fig:N-crit}
\end{figure}

In Figure~\ref{Fig:N-crit}, we present our results for the three Pop.~III stars considered in this work. The light-blue shaded area denotes the DM-proton cross sections for which DM does not thermalize. The regions where DM self-gravitate are shaded in magenta. Since the capture rate is larger for heavier stars, we observe that the self-gravitation region increases from the $20\Msun$ star to the $1000\Msun$ case. The orange areas depict the regions of the $m_\chi-\sigma_{\chi p}$ plane where gravitational collapse occurs, and the BH formed is able to accrete the entire star before the expected end of the star's life. Remarkably, this could occur in regions of the parameter space that have not been probed yet by direct detection experiments (shaded gray areas)~\cite{LZ:2024psa}, for the $20\Msun$ and $100\Msun$ stars. To the right of the orange areas, a BH may form, but accretion occurs at a slower rate such that no significant part of the star has been consumed before total depletion of helium from its core. For a heavier star, it takes longer for the BH to accrete the star's mass, which explains why the orange area shrinks for the larger stellar masses. In addition, less massive stars live longer, which means the BH has more time to accrete less stellar matter. It is worth noting that in the orange regions, at all times, the Schwarzschild radius of the BH is always larger than the size of the atoms being accreted, i.e., there are no quantum effects present in BH accretion for the stars considered~\cite{Giffin:2021kgb}. Notice that these results are obtained with a DM density, $\rho_\chi=5.3\,{\rm GeV/cm^3}$, much lower than the densities considered in previous works~\cite{Iocco:2012wk, Ilie:2019sjk,Croon:2023trk}.

Finally, we would like to comment on BH evaporation. There is an additional contribution to $\operatorname{d\!}\MBH/\operatorname{d\!}t$ from the emission of Hawking radiation by the BH. The new equation for the BH evolution is then
\begin{equation}\label{eq:bhevol}
    \frac{\operatorname{d\!}\MBH}{\operatorname{d\!}t}=\frac{\operatorname{d\!}\MBH}{\operatorname{d\!}t}\bigg|_{\rm Acc}-\frac{\operatorname{d\!}\MBH}{\operatorname{d\!}t}\bigg|_{\rm Evap}\,,
\end{equation} with the first term on the right hand side given by Eq.~\eqref{eq:BHaccrate}, and 
\begin{equation}
    \frac{\operatorname{d\!}\MBH}{\operatorname{d\!}t}\bigg|_{\rm Evap} = 5.34 \times 10^{16}\,f(\MBH)\, \left(\frac{{\rm kg}}{\MBH}\right)^2\,{\rm kg/s}\,,
\end{equation} where $f(\MBH)$ is a factor determined by the number of degrees of freedom available for emission, depending on the BH temperature. In our results, the evaporation rate is several orders of magnitude smaller than the accretion rate in the regions where the BH accretes the whole star. In all three stars we are considering, the evaporation and accretion rates are comparable only for BH masses $\MBH \gtrsim 10^{12}\,\Msun$. 

\section{Conclusions}
\label{sec:conclusions}

We have considered heavy dark matter (DM) accreted onto massive stars, using an improved treatment of multiple-scattering capture that incorporates realistic stellar structure and velocity distributions. Focusing on Population III (Pop.~III) stars, we employ stellar evolution simulations of the first stars to model the radially dependent density, escape velocity, and temperature profiles, as they change over time, and use them to evaluate DM capture, thermalization, and annihilation at different stages of the star's evolution.

Our findings show that radial variations in escape velocity and the suppression of the DM low-velocity tail in Eddington-inverted phase-space distributions, which are better suited to stars at close distances from the halo center, both play important roles in DM capture calculations. Previous work had considered only the early stage of the star, composed of hydrogen and helium. We find that previous estimates based on constant-density and escape-velocity models, both evaluated at the star's surface, and on Maxwell-Boltzmann velocity distributions overpredict capture rates by more than an order of magnitude. The late stage of the star, in which a metal-rich core forms, had not been considered in the literature for these stars. In this stage, capture becomes sensitive to scattering on multiple nuclear species. We extended the response-function formalism from Ref.~\cite{Bell:2024qmj} to include three targets and demonstrated that this contribution is essential for accurately modeling capture of heavy DM with $m_\chi \gtrsim 10^{9}\,\mathrm{GeV}$. 

Our analysis of DM thermalization, annihilation, and self-gravitation shows that, for a wide range of parameters, captured DM efficiently thermalizes within the stellar core. For annihilating DM, capture and annihilation rapidly reach equilibrium, preventing a significant buildup of DM. In contrast, for non-annihilating DM, the accumulated population may become self-gravitating and, in parts of parameter space, undergo gravitational collapse, potentially leading to black hole formation and stellar destruction before the end of the star’s lifetime.

Our results demonstrate that DM capture effects in the first stars, and in general in massive stars, are highly sensitive to stellar evolution, internal composition, and the surrounding halo environment. Accurate modeling of these effects is essential for deriving robust constraints on heavy DM from primordial stellar populations and for assessing the potential astrophysical impact of DM in the first generation of stars.

\acknowledgments
We thank Paolo Salucci for
helpful and insightful discussions. 
SR  was supported by the Fermi National Accelerator Laboratory (Fermilab), a U.S. Department of Energy, Office of Science, HEP User Facility. 
This work was performed in part at the Aspen Center for Physics, which is supported by National Science Foundation grant PHY-2210452. 
SR also acknowledges the CERN TH Department for its hospitality while this research was being carried out. 
WT is supported by the National Science Foundation under Grant No.~PHY-2310224. WT is grateful to the Physics Department at Northwestern University, the Galileo Galilei Institute for Theoretical Physics, and the Physics Department at IIT-Bombay for their hospitality during the final stages of this work. WT is also grateful to the Faculty Development Leave Program of Loyola University Chicago for providing the valuable time we needed to complete this work. 
GB was supported by the Australian Research Council through the ARC Centre of Excellence for Dark Matter Particle
Physics, CE200100008. 
This research was undertaken using the Computational Research, Engineering and Technology Environment (CREATE)~\cite{CREATEHPC} at King's College London. 

\appendix

\section{Nuclear response functions} \label{app:nuclear-response}
\subsection{Multi-ionic elements}
\label{app:nuclear-responsefunc}
In this section, we summarize the expressions from Ref.~\cite{Bell:2024qmj} for the multiple scattering response function. This treatment accounts for the radial dependence of the escape velocity and target number density, nuclear form factors, and gravitationally curved trajectories of DM.

To model multiple collisions between the DM particle and the target $i$, we define the probability density for the DM energy loss in terms of the recoil energy of the target  $E_R$:
\begin{equation}
f(E_R)=\frac{1}{\sigma_{\chi i}}\frac{d\sigma_{\chi i}}{dE_R}.
\end{equation}
The probability of losing at least $\delta E = m_\chi u_\chi^2/2$ in one collision is
\begin{equation}
{\cal F}_1(\delta E)=\int_{\delta E}^{\infty} dE_R f(E_R).
\end{equation}
Assuming that $d\sigma_{\chi i}/dE_R\sim e^{-E_R/E_0}$, the analogous probability after $N$ scatterings is given by
\begin{equation}
{\cal F}_N(\delta)=\frac{e^{-\delta}\delta^{N-1}}{(N-1)!}\,,
\qquad
\delta\equiv \delta E/E_0.
\end{equation}
The probability that the DM particle experiences $N$ scatterings, while traversing an optical depth $\tau_\chi$, is given by the Poissonian probability 
\begin{equation}
p_N(\tau_\chi)=e^{-\tau_\chi}\frac{\tau_\chi^N}{N!}.
\end{equation}

Summing over all $N$ collisions, we obtain the ``response function'' for multiple scattering
\begin{eqnarray}
    G(\tau_\chi,\delta) &\equiv& \sum_{N=1}^\infty p_{N-1}(\tau_\chi) {\cal F}_N(\delta) = \sum_{N=1}^\infty \frac{e^{-\tau_\chi}\tau_\chi^{N-1}}{(N-1)!}  \frac{e^{-\delta}\delta^{N-1}}{(N-1)!} \nonumber\\
    &=& e^{-\tau_\chi-\delta}I_0\left(2\sqrt{\tau_\chi\delta}\right),
\end{eqnarray}
which is the probability to lose at least an amount of energy  $\delta E$ after $N$ collisions, while traveling and optical depth $\tau_\chi$. 

Next, we  consider the two possible paths a DM particle can follow to scatter with a star's constituent, which yield the two optical depths~\cite{Bell:2020jou,Bell:2021fye}
\begin{eqnarray}
    \tau^{-}_{\chi,i}(r,y)&=& \int_{r}^{R_\star} \operatorname{d}\!x\frac{n_i(x) \sigma_{\chi\, i}(x)}{\sqrt{1-y^2 J^2_{\rm max}(r)/J^2_{\rm max}(x)}}\, , \\
    \tau^{+}_{\chi,i}(r,y)&=& 2\tau^{-}_{\chi,i}(r_{\rm min},y)-\tau^{-}_{\chi,i}(r,y)\, ,
\end{eqnarray} 
where $J_{\rm max}$ is the maximum value of the angular momentum of the particle, $J$, and $y\equiv~J/J_{\rm max}$. Averaging over the two possible paths and the angular momentum yields the final response function, $\tilde G(r,\delta)$.

 \subsection{The hydrogen response function}
 \label{app:H-responsefunc}
 \subsubsection{Exact result}
The form factor for a hydrogen nucleus is trivial; hence, the expression for the response function in the previous section cannot be applied to calculate the capture rate. Here, we derive a hydrogen response function for the first time. Our starting point is the definition of the probability density function for the DM energy loss, as given in Ref.~\cite{Bell:2024qmj}. We consider a differential cross-section of the form
 \begin{equation}\label{eq:prob0}
 f(E_R)\,=\,\frac{1}{\sigma_{p\chi}}\frac{\operatorname{d}\!\sigma_{p\chi}}{\operatorname{d}\!E_R}(E_R)=\frac{1}{E_R^\text{max}} \Theta\left(E_R^\text{max}-E_R\right),
 \end{equation}
 where $E_R^\text{max}$ is the maximum recoil energy. 
 We can define a new variable, $\delta\equiv E_R/E_R^\text{max}$, so that
 \begin{equation}\label{eq:prob1}
 f(\delta) = \Theta\left(1-\delta\right).
 \end{equation}
 Taking the Laplace transform of the previous expression for one to $N$ collisions, 
 \begin{eqnarray}
     \tilde{f}(s)&=& \frac{1-e^{-s}}{s},\qquad \tilde{\mathcal{F}}_1 = \frac{1-e^{-s}}{s}-\frac{e^{-s}(e^s-1-s)}{s^2},\\
     \tilde{\mathcal{F}}_N(s) &=& \left(\frac{1-e^{-s}}{s}-\frac{e^{-s}(e^s-1-s)}{s^2}\right)\left(\frac{1-e^{-s}}{s}\right)^{N-1} \, .
 \end{eqnarray}
 Inverting this relation gives $\mathcal{F}_N$. For $\delta<1$, the result is simply
 \begin{equation}
 \mathcal{F}_N(\delta)\,=\,\frac{\delta^{N-1}}{(N-1)!}\left(1-\frac{\delta}{N} \right) \Theta (1-\delta),\quad \delta<1. 
 \end{equation}
 This results in the following response function
 \begin{equation}
 \!G(\tau_\chi, \delta ) = \sum_{N=1}^\infty p_{N-1}(\tau_\chi) \mathcal{F}_N(\delta) = \Theta (1-\delta)  e^{-\tau_\chi} \left[I_0\left(2\sqrt{\tau_\chi \delta}\right) -\sqrt{\frac{\delta}{\tau_\chi}}\,I_1\left(2\sqrt{\tau_\chi \delta}\right) \right]. 
 \end{equation}
On the other hand, for $\delta>1$,
 \begin{equation}
 \mathcal{F}_N(\delta)=\sum_{m=0}^{\max\{[\delta+1],N-1\}} (-1)^{m+1}\frac{(\delta-m)^{N-1}(\delta-N)}{\Gamma(N+1-m)\Gamma(m+1)} \Theta (\delta-m)\, ,
 \end{equation}
 where $[\delta]$ indicates the integer part of $\delta$. This results in 
 \begin{eqnarray}
   G(\tau_\chi, \delta ) &=&  \sum_{m=0}^\infty \Theta (\delta-m)\frac{ (-1)^{m+1} e^{-\tau_\chi} \sqrt{\tau_\chi} }{\Gamma (m+1)}\left(\sqrt{\tau_\chi} \sqrt{\delta
    -m}\right)^{m-2}\\
     &\times&  \left((\delta -1) \sqrt{\delta -m}
    I_{m-1}\left(2 \sqrt{\tau_\chi} \sqrt{\delta -m}\right)+\sqrt{\tau_\chi}
    (m-\delta ) I_{m-2}\left(2 \sqrt{\tau_\chi} \sqrt{\delta
    -m}\right)\right)\, . \nonumber
 \end{eqnarray}
 For a fixed value of $\delta$, this is a finite series of exponentially oscillating terms of alternating sign, whose sum will cancel out to large precision. Therefore, this expression is highly unsuitable to be used for high $\delta$, where many such terms contribute and need to be accounted for to obtain a meaningful result. Hence, this expression is only suitable for moderate values of $\delta>1$.
 When $\delta\gg 1$, we can use the Central Limit Theorem (CLT) to derive an approximated result.

\subsubsection{High energy loss approximation for the cumulative response function}

We define the cumulative response
\begin{equation}
H(\tau_\chi,\delta) \;=\; \int_0^\delta G(\tau_\chi,u)\,\operatorname{d}\!u,
\end{equation}
where \(G(\tau_\chi,u)\) admits the Poisson-mixture representation
\begin{equation} \label{eq:Poisson-mix}
G(\tau_\chi,u) = e^{-\tau_\chi}\sum_{n=1}^{\infty}\frac{\tau_\chi^{\,n}}{n!}\bigl(F_n(u)-F_{n+1}(u)\bigr),
\end{equation}
and \(F_n\) denotes the cumulative distribution function (CDF) of the sum \(S_n=\sum_{i=1}^n X_i\) of \(n\) independent uniform \([0,1]\) random variables (the Irwin--Hall distribution).  Exchanging the sum and the integral gives
\begin{equation} \label{eq:PoissonH}
H(\tau_\chi,\delta) = e^{-\tau_\chi}\sum_{n=1}^{\infty}\frac{\tau_\chi^{\,n}}{n!}\,J_n(\delta), \qquad J_n(\delta) \equiv \int_0^\delta \bigl(F_n(u)-F_{n+1}(u)\bigr)\,\operatorname{d}\!u .\\
\end{equation}

The quantity \(J_n(\delta)\) can be written probabilistically as a difference of truncated expectations, namely
\begin{equation}
J_n(\delta) =\mathbb{E}\big[(\delta-S_n)_+\big] - \mathbb{E}\big[(\delta-S_{n+1})_+\big],
\end{equation}
where \((x)_+=\max\{x,0\}\).  For large \(n\), the CLT applies to \(S_n\), with 
\begin{equation}
\mu \;=\; \mathbb{E}[X_i] \;=\; \tfrac{1}{2}, \qquad
\sigma^2 \;=\; \operatorname{Var}(X_i) \;=\; \tfrac{1}{12},
\end{equation}
the local Gaussian approximation for the density of \(S_n\) is
\begin{equation} \label{eq:gauss_density}
p_{S_n}(x) \;\approx\; \frac{1}{\sqrt{2\pi n\sigma^2}}
\exp\!\Big(-\frac{(x-n\mu)^2}{2n\sigma^2}\Big).
\end{equation}

Using this Gaussian approximation, we obtain a closed-form expression for the truncated expectation, 
\begin{equation}
H(\tau_\chi,\delta) \;\approx\; e^{-\tau_\chi}\sum_{n=1}^{\infty}\frac{\tau_\chi^{\,n}}{n!}\,
\Big\{(\delta-\tfrac{n}{2})\,\Phi(z_n)+\sqrt{\tfrac{n}{12}}\,\varphi(z_n) 
- (\delta-\tfrac{n+1}{2})\,\Phi(z_{n+1}) -\sqrt{\tfrac{n+1}{12}}\,\varphi(z_{n+1})
\Big\} \;,
\end{equation} where 
\begin{equation}
\varphi(z) \equiv \frac{1}{\sqrt{2\pi}} e^{-z^2/2}, \qquad
\Phi(z) \equiv \int_{-\infty}^{z}\varphi(y)\,dy .
\end{equation}

The practical advantages of this representation are
\begin{enumerate}[label=(\roman*),itemsep=0ex]
\item Each $J_n(\delta)$  is bound and smooth in $\delta$.
\item The Poisson weights $e^{-\tau_\chi}\tau_\chi^{\,n}/n!$  ``cut off'' the sum, so the series converges rapidly. 
\item  $H(\tau_\chi,\infty)\to 1$ with uniformly small Gaussian approximation errors for large $n$. 
\end{enumerate}

For numerical implementations, the series will have to be truncated at some large value of $n=n_{\rm switch}$,
\begin{equation}
H(\tau_\chi,\delta) \approx e^{-\tau_\chi}\sum_{n=1}^{n_{\rm switch}} \frac{\tau_\chi^{\,n}}{n!} J_n(\delta).
\end{equation} 
This yields a numerically stable approximation for all \(\delta\), and the pointwise density \(G(\tau_\chi,\delta)\) is then recovered by numerical differentiation
\begin{equation}
G(\tau_\chi,\delta) \approx \frac{\partial}{\partial \delta} H(\tau_\chi,\delta),
\end{equation}
which is far more stable than approximating \(G\) term by term.

This expression still has one issue, as it must be summed over all terms up to $n=n_{\rm switch}\sim\tau_\chi$ (similar to previous results). This turns out to be a problem when $\log(n_{\rm switch})\gg 1$, as it requires the sum of a very large number of terms. Next, we explore a solution to this issue.

\subsubsection{Saddle point asymptotics for the response function}

For large parameters, when $\tau_\chi\sim 2\delta$, the dominant contribution to the sum in Eq.~\eqref{eq:Poisson-mix} comes from a single \(n\)-region \(n\approx n^*\) determined by a saddle-point condition. To construct the asymptotic behavior, we approximate the sum argument by
\begin{equation}
a(n)\equiv \frac{1}{2} e^{-\tau_\chi}\frac{\tau_\chi^{\,n}}{n!}\,p_{S_n}(\delta),
\end{equation}
where \(p_{S_n}\) is given in \eqref{eq:gauss_density}. 
We define the log-amplitude (dropping additive constants independent of \(n\)) as
\begin{equation}
\mathcal{L}(n)\equiv\ln\left(2a(n)\right)
\approx -\tau_\chi + n\ln(\tau_\chi) -\ln(\Gamma(n+1)) -\frac{(\delta-n\mu)^2}{2n\sigma^2} -\tfrac12\ln(2\pi n\sigma^2).
\end{equation}

The saddle point \(n^*\) satisfies \(\mathcal{L}'(n^*)=0\), which (keeping dominant powers) gives the transcendental equation
\begin{equation}
\ln\left(\frac{\tau_\chi}{n^*}\right) \;+\; \frac{\mu(\delta-n^*\mu)}{n^*\sigma^2} \;+\; \frac{(\delta-n^*\mu)^2}{2(n^*)^2\sigma^2} \;\approx\; 0.
\label{eq:SPgeneral}
\end{equation}

When \(\tau_\chi\) differs from \(2\delta\) but remains in the saddle point regime, we solve Eq.~\eqref{eq:SPgeneral} for \(n^*\) and evaluate the Laplace approximation
\begin{equation}
G(\tau_\chi,\delta)\;\approx\; \frac{1}{2}\,e^{-\tau_\chi}\frac{\tau_\chi^{\,n^*}}{\Gamma(n^*+1)}\,
\frac{1}{\sqrt{2\pi n^*\sigma^2}}\;
\exp\!\Big(-\frac{(\delta-n^*\mu)^2}{2n^*\sigma^2}\Big)\;
\sqrt{\frac{2\pi}{-\mathcal{L}''(n^*)}}\,,
\label{eq:G_general}
\end{equation}
with \(\mu=\tfrac12\), \(\sigma^2=\tfrac{1}{12}\) and \(\mathcal{L}''(n)\) obtained by differentiating \(\mathcal{L}(n)\).

The approximation \eqref{eq:G_general} is uniformly accurate when the dominant \(n^*\) is large, and the saddle is sharp, and the relative error is \(O(1/\sqrt{n^*})\).

The previous approximation is valid for large $\delta$, but for moderate values (e.g.\ $\delta\sim 30$) it can produce an overall multiplicative bias of order unity. 
A more accurate treatment starts from the identity
\[
F_n(\delta)-F_{n+1}(\delta)
= \int_0^1 (1-y)\, p_{S_n}(\delta-y)\, \operatorname{d}\!y,
\]
and expands the integrand in powers of $y$. One finds
\begin{equation}
\label{eq:Fn-diff-expansion}
F_n(\delta)-F_{n+1}(\delta)
= \tfrac{1}{2}\,p_{S_n}(\delta)
-\tfrac{1}{6}\,p_{S_n}'(\delta)
+\tfrac{1}{24}\,p_{S_n}''(\delta)+\cdots.
\end{equation}

Equation~\eqref{eq:Fn-diff-expansion} then takes the compact form
\begin{equation}
\label{eq:M-corr}
F_n(\delta)-F_{n+1}(\delta)\;\approx\;
M(n,\delta)\, p_{S_n}(\delta),
\end{equation}
with the correction factor
\begin{equation}
\label{eq:Mcorr-def}
M(n,\delta) \;=\; \tfrac{1}{2}
+ \frac{X}{6A}
+ \frac{1}{24}\!\left(\frac{X^2}{A^2}-\frac{1}{A}\right)
+ \cdots.
\end{equation}

Finally, evaluating at the saddle $n=n^\ast$, and including the usual Laplace prefactor, the corrected saddle point approximation reads
\begin{equation}
\label{eq:G-corrected}
G_{\rm corrected}(\tau_\chi,\delta)\;\approx\;
2M(n^\ast,\delta)\;
G(\tau_\chi,\delta)\, .
\end{equation}
The leading factor $1/2$ is thus replaced by the improved multiplier
$M(n^\ast,\delta)$. 
This removes much of the bias observed in
comparisons between the raw saddle point formula and the exact summation.

\section{Nuclear form factors}\label{app:form-factors}

The differential scattering cross section between the DM particle and a nucleus $i$ for interaction in Eq.~\eqref{eq:eff_L} is given by
\begin{equation}
    \frac{\operatorname{d}\!\sigma_{i \chi}}{\operatorname{d}\! \cos\theta_\text{cm}}= \frac{1}{2\pi\Lambda^4}\frac{m_\chi^2m_i^2}{(m_\chi+m_i^2)m_N^2}\sum_{N,N'} c_N^S c_{N'}^S F^{N,N'}_i(E_R),
    \label{eq:diffxsec}
\end{equation} 
where $\theta\text{cm}$ is the center of mass angle, $E_R$ is the recoil energy, $N,N'=p,n$ represent nucleons. 
The DM-nucleon scattering cross section $\sigma_{\chi N}$ is defined as 
\begin{equation}
    \sigma_{\chi N}\,=\,\left(\frac{c_N^S}{\Lambda^2}\right)^2 \frac{\mu_{\chi N}^2}{\pi},
\end{equation} where $\mu_{\chi N}$ is the DM-nucleon reduced mass, $c_p^S=0.00162622$ and $c_n^S=0.00164864$, are the proton and neutron hadronic matrix elements of the scalar effective operator, respectively~\cite{Bell:2021fye}. 
The relevant nuclear form factors in Eq.~\eqref{eq:diffxsec} reduce to the $F_i(E_R)$ values given in Table~\ref{tab:formfactors}.

\begin{table}[t]
    \centering
    \begin{tabular}{|l|c|c|}
    \hline
       Element  &  $E_1^i ({\rm MeV})$ & Form factor $F_i(E_R)$ \\
     \hline  
       Helium  &  $4.36$ & $ 4\pi\times0.31831\, \exp\left({-\frac{E_R}{E^{\rm He}_1}}\right)$ \\
       Carbon & $1.78$ &  $ 4\pi\times0.585882\, \exp\left({-\frac{E_R}{E^{\rm C}_1}}\right)\left(2.25-\frac{E_R}{2E^{\rm C}_1}\right)^2$\\
       Oxygen & $1.10$ & $ 4\pi\times0.000032628\, \exp\left({-\frac{E_R}{E^{\rm O}_1}}\right)\left(395.084-200.042\frac{E_R}{2E^{\rm O}_1}+\frac{E_R^2}{4(E^{\rm O}_1)^2} \right)^2$\\
       Neon &  $0.62$ & $4\pi\times0.0431723\, \exp\left({-\frac{E_R}{E^{\rm Ne}_1}}\right)\left(13.577-9.05108\frac{E_R}{2E^{\rm Ne}_1}+\frac{E_R^2}{4(E^{\rm Ne}_1)^2} \right)^2$ \\
     \hline  
    \end{tabular}
    \caption{Nuclear form factors, taken from Ref.~\cite{Catena:2015uha}.}
    \label{tab:formfactors}
\end{table}

\section{The Eddington inversion method}\label{app:Eddington}
The Eddington inversion method allows us to obtain the phase space distribution function (PSDF) for a system in dynamical equilibrium~\cite{Eddington:1916,Binney:2008}. In a spherically symmetric system, in the presence of a gravitational potential $\Phi(r)$ due to a mass density $\rho(r)$, we can define the relative potential 
\begin{align}
    \Psi(r) &= \Phi(R_{\rm max})-\Phi(r)= \int_{r}^{R_{\rm max}} \operatorname{d}\! r' \frac{G\,m(r')}{r'^{2}}\, , \\ m(r)&=\, 4\pi\int_0^r \operatorname{d}\! r' r'^2 \rho(r'),
\end{align} 
where $R_{\rm max}$ is the radius of the boundary of the system. 
The Eddington formula yields the PSDF
\begin{equation}
    F(\mathcal{E})\,=\,\frac{1}{\sqrt{8} \,\pi^2} \left[\frac{1}{\sqrt{\mathcal{E}}}\left (\frac{\operatorname{d}\!\rho}{\operatorname{d}\! \Psi}\right)_{\Psi=0} +\int_0^\mathcal{E} \frac{\operatorname{d}^2\!\rho}{\operatorname{d}\! \Psi^2} \frac{\operatorname{d}\!\Psi}{\sqrt{\mathcal{E}- \Psi}} \right],
\end{equation} 
where $\mathcal{E}$ is the relative energy per unit mass, $\mathcal{E}=\Psi(r)-u^2/2$, with $u$ the velocity modulus. 

In the frame of a star with speed $v_\star$, the normalized DM velocity distribution function is given by~\cite{Lopes:2020dau} 
\begin{equation}
    f_{\chi\star}(r,u_\chi)\,=\,2\pi\,u_\chi^2 \int_0^\pi \operatorname{d}\! \theta\, \sin \theta\,\frac{F(\mathcal{E})}{\rho(r)}\, ,
\end{equation} with $\mathcal{E}\,=\,\Psi(r)- \frac{1}{2}(u_\chi^2+v_\star^2+2u_\chi v_\star\,\cos \theta)$, where $\theta$ is the angle between $\vec{u}_\chi$ and $\vec{v}_\star$.

\section{Capture rate scaling for heavy dark matter} \label{sec:scaling}

At large DM mass $m_\chi\gg m_i$, the DM-target differential cross section $\frac{\operatorname{d}\!\sigma_{\chi i}}{\operatorname{d}\!E_R}$ and $\tau_\chi^\pm$ become independent of $m_\chi$. 
Taking the limit $m_\chi\gg m_i$ of the ratio $C/C_\text{geom}$ for heavy DM and scattering with a single target species,
\begin{equation}
   \frac{C}{C_\text{geom}} \propto \int_0^{R_\star} \operatorname{d}\!r \, 4\pi r^2  n_T(r)\sigma_{\chi i}(r)v_\text{esc}^2(r)\int_0^1 \frac{ydy}{\sqrt{1-y^2}}\int_0^\infty \operatorname{d}\!u_\chi \frac{f_{\chi\star}(u_\chi)}{u_\chi}G\left(\tau_\chi(r,y),\frac{m_\chi u_\chi^2}{2E_0}\right),
   \label{eq:Cratio}
\end{equation}
the only remaining dependence on $m_\chi$ is encoded in the second argument of the response function $G$, from which we can obtain the scaling at large mass. We also notice that the integral over the velocity distribution does not extend up to infinity, but is limited by the DM escape velocity from the halo $u_\text{max}=v_\text{esc}^\text{halo}(r_\star)$. 
Now, if 
\begin{equation}
   \delta(u_\text{max}) = \frac{m_\chi u_\text{max}^2}{2E_0} \ll 1 \rightarrow  m_\chi \ll  \frac{2E_0}{u_\text{max}^2} \equiv m^*,  \label{eq:mstardef}
\end{equation} where $m^*$ is the DM mass for which capture requires multiple scatterings, then we can approximate the response function as
\begin{equation}
   G(\tau_\chi,\delta)\rightarrow G(\tau_\chi,0) = e^{-\tau_\chi}.
\end{equation}
This means that for $m_i \ll m_\chi \ll m^*$,  the ratio $C/C_\text{geom}$ is independent of $m_\chi$.

Let's now consider the case $m_\chi \gg m^*$ instead. 
If the cross-section is small, we can approximate
$G(\tau_\chi,\delta)\,\sim\, e^{-\delta}$. 
With our assumption on $m_\chi$, written as $u_\text{max}\ll v_\text{esc}$, and if $v_\text{esc}$ is of a similar order of the other typical speed scales entering in $f_{\chi\star}$, one can approximate $f_{\chi\star}$ with its expansion around $u_\chi=0$. 
Then the integral over $u_\chi$ becomes 
\begin{equation}
   \int_0^\infty \operatorname{d}\!u_\chi \frac{f_{\chi\star}(u_\chi)}{u_\chi} e^{-\frac{m_\chi u_\chi^2}{2E_0}} 
   \,\sim\, f'_{\chi\star}\int_0^{u_\text{max}} \operatorname{d}\!u_\chi u_\chi e^{-\frac{m_\chi u_\chi^2}{2E_0}},
   \label{eq:fuint}
\end{equation}
where $f'_{\chi\star}$ is the coefficient of the expansion. 
Since the integrand in the latter expression rapidly approaches $0$ for large $u_\chi$, we can extend the integration interval to $[0,\infty)$. 
For a Maxwell-Boltzmann distribution, this can be explicitly verified. Performing a change of variable, $w^2=\frac{m_\chi u_\chi^2}{2E_0}$ and using that $\sigma_{\chi\,i} (v_\text{esc})=\frac{k(v_\text{esc})}{\Lambda^4}  \,\propto \,\Lambda^{-4}$, 
where $k$ is a function of the escape velocity and is independent of $m_\chi$, 
we obtain
\begin{equation}
   \frac{C}{C_\text{geom}} \propto \int_0^{R_\star} \operatorname{d}\!r \, 4\pi r^2  n_T(r)\frac{k(v_\text{esc})}{\Lambda^4}v_\text{esc}^2(r)\int_0^\infty \frac{\operatorname{d}\!w\, w}{m_\chi} e^{-w^2}. 
\end{equation}
This quantity is invariant under the rescaling $m_\chi\rightarrow \alpha \, m_\chi$ and $\Lambda\rightarrow \Lambda \,\alpha^{-1/4}$, where $\alpha$ is a constant.

A very similar reasoning applies when the cross-section is not small, i.e., when the opacity is large. In this case, we cannot approximate the response function as a simple exponential. However, for large $\tau_\chi$ the response function acts as a Dirac delta function
\begin{equation}
   G(\tau_\chi,\delta) \sim \delta\left(\tau_\chi-\delta\right), 
\end{equation}
and the ratio $\frac{C}{C_\text{geom}}$ is given by
\begin{equation}
   \frac{C}{C_\text{geom}} \sim \frac{\int_0^{\tau^\text{max}_\chi}\operatorname{d}\!\tau_\chi\int_0^\infty \operatorname{d}\!u_\chi \frac{f_{\chi\star}(u_\chi)}{u_\chi}G(\tau_\chi,\frac{m_\chi u_\chi^2}{2E_0})}{\int_0^\infty \operatorname{d}\!u_\chi \frac{f_{\chi\star}(u_\chi)}{u_\chi}}. 
   \label{eq:Cratiolargexsec}
   \end{equation}
If we rescale $\operatorname{d}\!\sigma_{\chi i}\rightarrow\alpha \operatorname{d}\!\sigma_{\chi i}$ and $m_\chi\rightarrow\alpha \,m_\chi$, 
Eq.~\eqref{eq:Cratiolargexsec} transforms as
\begin{eqnarray}
   \frac{C}{C_\text{geom}}
  &=& \frac{\int_0^{\tau^\text{max}_\chi}\operatorname{d}\!\tau_\chi\int_0^\infty \operatorname{d}\!u_\chi \frac{f_\chi\star(u_\chi)}{u_\chi}\delta(\tau_\chi-\frac{m_\chi u_\chi^2}{2E_0})}{\int_0^\infty \operatorname{d}\!u_\chi \frac{f_{\chi\star}(u_\chi)}{u_\chi}} \nonumber\\
  &\rightarrow&\frac{\int_0^{\tau^\text{max}_\chi} \alpha \operatorname{d}\!\tau_\chi\int_0^\infty \operatorname{d}\!u_\chi \frac{f_{\chi\star}(u_\chi)}{u_\chi}\delta(\alpha(\tau_\chi-\frac{m_\chi u_\chi^2}{2E_0}))}{\int_0^\infty \operatorname{d}\!u_\chi \frac{f_{\chi\star}(u_\chi)}{u_\chi}}, 
\end{eqnarray}
The integral of the Dirac delta function over $\tau_\chi$ yields $1$ for
\begin{equation}
   \tau^\text{max}_\chi>\frac{m_\chi u_\chi^2}{2E_0}, 
\end{equation} 
and $0$, otherwise. The dependence on $\alpha$ is dropped due to the properties of the Dirac delta function. The geometric limit is reached when
\begin{equation}
   \tau^\text{max}_\chi\sim\frac{m_\chi u_\text{max}^2}{2E_0}.
\end{equation}
Notice that this condition is also invariant under rescaling.

\bibliographystyle{JHEP}
\bibliography{referencesPopIII}
\end{document}